\pdfoutput=1

\documentclass[11pt,twoside,a4paper,cmspaper,final,collab]{cms-tdr}
\def\svnVersion{322689}\def\cmsCernNo{2013-037}\def\cmsCernDate{\today}\def\cmsMessage{Published in the Journal of High Energy Physics as \href{http://dx.doi.org/10.1007/JHEP01(2016)166}{\doi{10.1007/JHEP01(2016)166.}}}

\begin{document}\cmsNoteHeader{B2G-14-005}

\hyphenation{had-ron-i-za-tion}
\hyphenation{cal-or-i-me-ter}
\hyphenation{de-vices}
\RCS$Revision: 316123 $
\RCS$HeadURL: svn+ssh://svn.cern.ch/reps/tdr2/papers/B2G-14-005/trunk/B2G-14-005.tex $
\RCS$Id: B2G-14-005.tex 316123 2015-12-26 15:24:34Z alverson $

\newcommand{\bs}{\ensuremath{\PQb^*}\xspace}
\newcommand{\bsL}{\ensuremath{\PQb^*_L}\xspace}
\newcommand{\bsR}{\ensuremath{\PQb^*_R}\xspace}
\newlength\cmsFigWidth
\ifthenelse{\boolean{cms@external}}{\setlength\cmsFigWidth{0.85\columnwidth}}{\setlength\cmsFigWidth{0.4\textwidth}}
\ifthenelse{\boolean{cms@external}}{\providecommand{\cmsLeft}{top}}{\providecommand{\cmsLeft}{left}}
\ifthenelse{\boolean{cms@external}}{\providecommand{\cmsRight}{bottom}}{\providecommand{\cmsRight}{right}}
\newcolumntype{x}[1]{D{,}{\,\pm\,}{#1}}
\newcolumntype{P}{D{p}{\%}{-1}}

\cmsNoteHeader{B2G-14-005}

\title{Search for the production of an excited bottom quark decaying to \texorpdfstring{ $\PQt \PW$ }{tW} in proton-proton collisions at \texorpdfstring{$\sqrt{s} = 8\TeV$}{sqrt(s) = 8 TeV}}

\date{\today}

\abstract{A search is presented for a singly produced excited bottom quark ($\bs$) decaying to a top quark and a $\PW$ boson in the all-hadronic, lepton+jets, and dilepton final states in proton-proton collisions at $\sqrt{s} = 8\TeV$ recorded by the CMS experiment at the CERN LHC. Data corresponding to an integrated luminosity of 19.7\fbinv are used. No significant excess of events is observed with respect to standard model expectations. We set limits at 95\% confidence on the product of the $\bs$ quark production cross section and its branching fraction to $\PQt \PW$. The cross section limits are interpreted for scenarios including left-handed, right-handed, and vector-like couplings of the $\bs$ quark and are presented in the two-dimensional coupling plane based on the production and decay coupling constants. The masses of the left-handed, right-handed, and vector-like $\bs$ quark states are excluded at 95\% confidence below 1390, 1430, and 1530\GeV, respectively, for benchmark couplings. This analysis gives the most stringent limits on the mass of the $\bs$ quark to date.
}

\hypersetup{%
pdfauthor={CMS Collaboration},%
pdftitle={Search for the production of an excited bottom quark decaying to tW in proton-proton collisions at sqrt(s) = 8 TeV},%
pdfsubject={CMS},%
pdfkeywords={CMS, physics, excited bottom quark, tW}}

\maketitle

\section{Introduction} \label{sec:intro}

Following the discovery of a Higgs boson~\cite{Aad:2012tfa,HiggsDiscovery,Chatrchyan:2013lba},
the standard model (SM) may be complete.
However, there are phenomena such as baryon asymmetry, neutrino mass,
and dark matter, questions of naturalness, and hierarchy problems for which the SM offers no explanation.
Various theories with new physics beyond the SM exist that address these problems,
including a variety of models that predict the existence of excited
quarks, such as
Randall--Sundrum models~\cite{Cheung:2007bu,Fitzpatrick:2007sa} and models
with a heavy gluon
partner~\cite{Bini:2011zb,Vignaroli:2012sf,Vignaroli:2012si}.
Searches for excited quarks have been performed at the
CERN LHC~\cite{ATLAS:2012pu,Chatrchyan:2013qha,Aad:2013rna} and elsewhere~\cite{pdg}.
These searches
focus on the strong and electroweak interactions of the excited quark with the
SM up- or down-type quarks. This paper reports on a search by the CMS Collaboration,
using the tW decay mode, for an excited third-generation bottom quark
($\bs$), which preferentially couples to the
third-generation SM quarks.
A previous search in the same channel by the ATLAS Collaboration
resulted in a lower limit on
the $\bs$ quark mass of about 1\TeV~\cite{Aad:2013rna}.
A search for a $\bs$ quark has also been performed in the $\Pg \PQb$ decay
mode by CMS~\cite{Chatrchyan:2013qha} resulting in an exclusion region between 1.2 and 1.6\TeV,
assuming a branching fraction of 100\% for \bs decaying to $\Pg \PQb$.

\begin{figure}[!h]
  \begin{center}
    \includegraphics[width=0.49\textwidth]{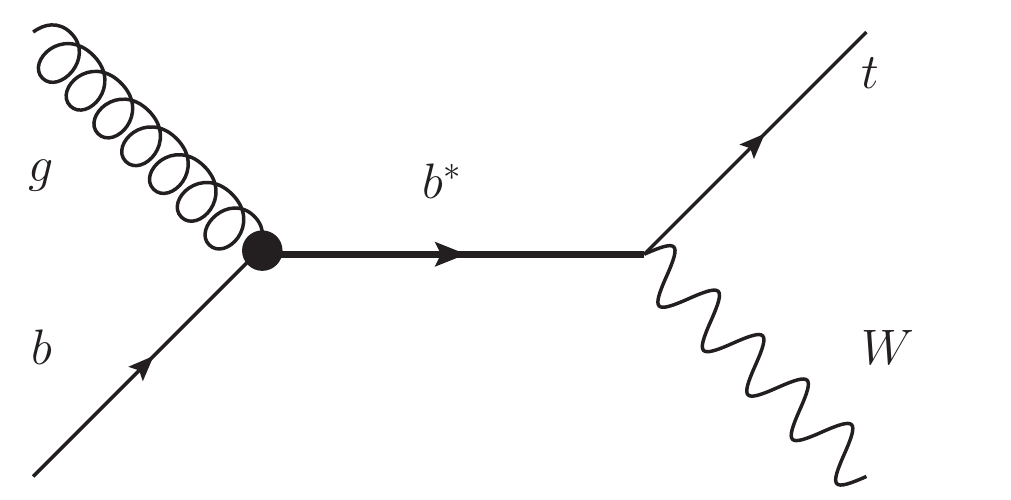}
   \caption{Leading-order Feynman diagram contributing to
     $ \Pg \PQb \to \bs \to \PQt \PW$.}
    \label{fig:bsFeynman}
  \end{center}
\end{figure}

At the LHC, a $\bs$ quark can be produced
in a gluon and a bottom quark interaction as shown in
Fig.~\ref{fig:bsFeynman}. This interaction is
described by the effective Lagrangian:

\begin{equation}
  {\mathcal L} = \frac{g_s}{2\,\Lambda} G_{\mu\nu}\,
          {\PAQb}\, \sigma^{\mu\nu}
          \biggl(\kappa^{\PQb}_L P_L + \kappa^{\PQb}_R P_R\biggr) \bs \xspace
          + \text{Hermitian conjugate (h.c.)},
\label{eq:productionL}
\end{equation}

where $g_s$ is the strong coupling, $G_{\mu\nu}$
is the gauge field tensor of the gluon, and $\Lambda$~\cite{PhysRevD.42.815} is the scale of compositeness,
which is chosen to be the mass of the $\bs$ quark. The quantities $P_L$ and
$P_R$ are the chiral projection operators and $\kappa^{\PQb}_L$ and
$\kappa^{\PQb}_R$ are the corresponding relative coupling strengths.
The branching fractions of $\bs$ quark decays are reported in Ref.~\cite{Nutter:2012an}.
Possible $\bs$ quark decay modes include $\Pg \PQb$,
$\PQb \PZ$, $\PQb \PH$, and $\PQt \PW$. The branching fraction of the
$\bs \to \PQt\PW$ process
 increases as a function of $\bs$ quark mass and becomes
the largest for $m_{\bs} > 400$\GeV, reaching a plateau at
almost 40\% of the total $\bs$ quark decay width.

The decay of interest in this analysis proceeds through the weak
interaction as is described by the
Lagrangian:

\begin{equation}
 {\mathcal L} = \frac{g_2}{\sqrt{2}} \,\PW^+_{\mu} \,\PAQt~\gamma^{\mu}\,
                  \bigl(g_L P_L + g_R P_R \bigr) \,\bs \xspace + \mathrm{h.c.},
   \label{eq:decayL}
\end{equation}

where $g_2$ is the weak coupling, and $g_L$ and $g_R$ are the relative
coupling strengths of the $\PW$ boson to the left- and right-handed $\bs$ quark, respectively.

This analysis searches for a singly produced $\bs$ decaying to a top quark and a $\PW$ boson.
Since there are both left- and right-handed
operators in the production and decay interaction Lagrangians, the
$\bs$ quark could have generic couplings. We consider the benchmark cases of
a purely left-handed \bs ($\bsL$) quark
 with $g_L$ = 1, $\kappa^{\PQb}_L$ = 1, $g_R$ = 0,
$\kappa^{\PQb}_R$ = 0, a purely right-handed \bs ($\bsR$) quark
 with $g_L$ = 0, $\kappa^{\PQb}_L$ = 0, $g_R$ = 1, $\kappa^{\PQb}_R$ = 1,
and a vector-like \bs quark with $g_L$ = 1, $\kappa^{\PQb}_L$
= 1, $g_R$ = 1, $\kappa^{\PQb}_R$ = 1.

The analysis is performed in three different channels distinguished by
the number of leptons (electrons and muons) appearing
in the $\bs \to \PQt \PW \to \PQb \PW \PW $ decay.
The all-hadronic channel has two jets: one from a boosted top quark
and the other from the boosted $\PW$ boson. As the Lorentz boosts of the
top quark and $\PW$ boson increase, the angular distance between their direct decay
products decreases, leading to only two resolvable jets.
The lepton+jets channel has one lepton, one $\PQb$ jet, two
light-flavor ($\PQu$-, $\PQd$-, $\PQs$-quark) or gluon jets, and significant
transverse momentum ($\pt$) imbalance.
The dilepton channel has two leptons,
at least one jet, and significant $\pt$ imbalance.

\section{The CMS detector}

The central feature of the CMS apparatus is a superconducting solenoid
of 6\unit{m} internal diameter, providing a magnetic field of
3.8\unit{T}. Within the solenoid volume are a silicon
pixel and strip tracker, a lead tungstate crystal electromagnetic
calorimeter (ECAL), and a brass and scintillator hadron calorimeter,
each composed of a barrel and two endcap sections.
Forward calorimeters extend the pseudorapidity coverage provided by
the barrel and endcap detectors. Muons are measured in gas-ionization
detectors embedded in the steel flux-return yoke outside the solenoid.
A detailed description of the CMS detector, together with a definition
of the coordinate system used and the relevant kinematic variables,
can be found in Ref.~\cite{Chatrchyan:2008zzk}.

\section{Signal and background simulations}
\label{samples}

The simulation of $\bs$ quark production and decay is performed with \MADGRAPH~5.1.5.12~\cite{madgraph5}
based on the Lagrangian in the $\bs$ quark model~\cite{Nutter:2012an}, and uses the {\sc CTEQ6L1}
parton distribution functions (PDF) set~\cite{CTEQ6L1}.
The renormalization and factorization scales are set to the $\bs$ quark mass.
The $\bs$ quark is forced to decay to tW, with the top quark
subsequently decaying into bW. The simulated samples are produced
for $\bs$ quark masses ranging from 800 to 2000\GeV, in steps of 100\GeV.
Left-handed and
right-handed $\bs$ quark samples are generated. The vector-like
$\bs$ quark samples are the sum of the right- and
left-handed samples.
The values for the $\bs$ quark production cross section times $ \bs \PQt \PW$  branching fraction in proton-proton collisions at a center-of-mass
energy of 8\TeV are listed in Table~\ref{bstarCrossSection}.

\begin{table}[htp]
\begin{center}
 \caption{
Estimates of the total cross section for $\Pg \PQb \to \bs $ at a center of mass energy
of 8\TeV times the branching fraction for  $\bs \to \PQt \PW$
for $\bs$ quark masses from 800 to 2000\GeV.
The values are identical for left-handed and right-handed quark hypotheses.
The uncertainties are determined by
varying the factorization ($\mu_{F}$) and renormalization ($\mu_{R}$) scales simultaneously
by a factor of 0.5 or 2 of their nominal value. The estimated cross section of a $\bs$ quark with vector-like
coupling is twice as large at each mass point as the value shown.}
   \begin{tabular}{cc|cc}
  \hline

    $\bs$ quark mass & $\sigma_{ \Pg \PQb \to \bs \to \PQt \PW} $ &  $\bs$ quark mass & $\sigma_{ \Pg \PQb \to \bs \to \PQt \PW }$ \\

    [\GeVns{}] & [pb] & [\GeVns{}] & [pb] \\
  \hline
    800  & 2.98    $\pm$ 0.39 & 1500 & 0.040 $\pm$ 0.006 \\
    900  & 1.45    $\pm$ 0.20 & 1600 & 0.024 $\pm$ 0.004 \\
    1000 & 0.74    $\pm$ 0.10 & 1700 & 0.014 $\pm$ 0.002 \\
    1100 & 0.39    $\pm$ 0.06 & 1800 & 0.009 $\pm$ 0.001 \\
    1200 & 0.21    $\pm$ 0.03 & 1900 & 0.005 $\pm$ 0.001 \\
    1300 & 0.12    $\pm$ 0.02 & 2000 & 0.003 $\pm$ 0.001\\
    1400 & 0.07    $\pm$ 0.01 &  &  \\
  \hline
   \end{tabular}
\label{bstarCrossSection}
\end{center}
\end{table}

Several simulated background samples are used.
The samples for $t$-channel, $ \PQt \PW$-channel, and $s$-channel production
of single top quarks, and the $\ttbar$ sample
are generated using the \POWHEG 1.0 event generator~\cite{Nason:2004rx,Frixione:2007vw,Alioli:2010xd} with the CT10
PDF set~\cite{PDF:CTEQ10}. A
next-to-next-to-leading-order (NNLO) cross section of 245.8\unit{pb} is used
for the $\ttbar$ sample~\cite{Czakon:2013goa}.
The total prediction is normalized to
the next-to-next-to-leading-log values of 87.1, 22.2, and
5.55\unit{pb} for the $t$-, $\PQt \PW$-, and $s$-channels, respectively~\cite{Kidonakis:2012db}.

The Drell--Yan sample (denoted as $\PZ$+jets in the following) with the invariant mass of two
leptons being greater than 50\GeV , and the $\PW$ inclusive sample ($\PW$+jets)
are generated using \MADGRAPH with the {\sc CTEQ6L1} PDF set.
The NNLO cross sections of 3500\unit{pb} and 36700\unit{pb} are
used for the $\PZ$+jets and $\PW$+jets normalization, respectively~\cite{Gavin:2010az}.

The diboson ($\PW \PW $, $\PW \PZ $, and $\PZ \PZ $) background samples are generated
inclusively
using \PYTHIA 6.426~\cite{Sjostrand:2006za} with the {\sc CTEQ6L1} PDF set
and normalized to a NNLO cross section of 57.1,
32.3, and 8.26\unit{pb}, respectively, calculated from MCFM 6.6~\cite{Campbell:2010ff}.

All of the samples are then interfaced to \PYTHIA for parton showering and
hadronization, based on the Z2* tune~\cite{Z2S}.
The generated samples are then passed to the CMS detector simulation
based on \GEANTfour~\cite{Agostinelli:2002hh}, with alignment and calibration determined
from data or dedicated calibration samples.
The average number of pileup interactions (additional inelastic
proton-proton collisions within the same bunch crossing)
is observed
to be approximately 20 for the data recorded in 2012.
Proton-proton collisions are added to simulated signal and background
events so that the distribution of reconstructed primary vertices agrees
with what is observed in data.

\section{Trigger, event quality, and object selection}
\label{objSel}

At least one reconstructed primary vertex that is associated with at
least four reconstructed tracks~\cite{TRK-11-001} is required to
be present in the event.

Events that are due to beam halo, poor calibration, and malfunctioning detector electronics
are rejected.
The particle-flow (PF) algorithm~\cite{CMS-PAS-PFT-09-001} is used for
both data and simulated events to reconstruct physics objects such as electrons, muons, and
charged and neutral hadrons.

Electron candidates are reconstructed within the
range of pseudorapidity $|\eta|$ $<2.5$
using the energy clusters in the ECAL~\cite{Khachatryan:2015hwa}. The clusters are
associated with charged-particle tracks reconstructed in the tracking
detector. The absolute value of the electron candidate transverse impact parameter should be
smaller than 0.02\unit{cm}. Identified electrons from photon conversions are
vetoed. The relative isolation requires $I_{\rm rel} < 0.1$, where
$I_{\rm rel}$ is the ratio of the sum of the $\pt$
of other particles around the electron candidate to the $\pt$ of the
electron candidate.
The $\pt$ summation is over the charged hadrons, photons, and
neutral hadrons,
in a cone size of $\Delta{R}=\sqrt{\smash[b]{{(\Delta\eta)}^2+{(\Delta\phi)}^2}}<0.3$,
where $\phi$ is the azimuthal angle in radians.
The estimated contribution from pileup is removed from the sum on an event-by-event basis~\cite{CMS-PAS-JME-13-005}.
Electron candidates with clusters
in the transition region between barrel and endcap
($1.4442<|\eta|<1.5660$) are removed
since the electron reconstruction is not optimal in this region. The $\pt$ of
the electron candidate is required to be larger than 30\GeV in the dilepton
channel and 130\GeV for the lepton+jets channel.

Muon candidates are reconstructed within $|\eta|<2.4$ by combining the
information from the muon detectors and the inner tracking detectors~\cite{Chatrchyan:2013sba}.
For the muon selection used in the lepton+jets channel, a requirement of $|\eta| < 2.1$ is imposed,
to match the coverage of the single muon trigger.
The candidate's trajectory fit has to satisfy $\chi^{2}/n < 10$
(where $n$ is the number of degrees of freedom in the fit),
have at least one hit in the muon detectors, and have more than five
hits in the silicon tracker, of which at least one should be
in the pixel detector. The absolute value of the muon candidate transverse impact
parameter should be smaller than 0.02\unit{cm}.
In order to suppress the small background due to cosmic ray muons,
the absolute value of the muon candidate longitudinal impact parameter
must be less than 0.5\unit{cm}. Isolated muons are
selected by the requirement $I_{\rm rel} < 0.12$ in a cone size of
$\Delta{R}<0.4$ around the muon candidate. The $\pt$ of
the muon candidate has the same threshold as the electron candidate.

The events are divided into all-hadronic, lepton+jets, and dilepton
channels based on the number of leptons (0, 1, or 2 leptons).
To suppress possible overlap between lepton+jets and dilepton  channels,
events with additional electrons (muons) with
$\pt>20\GeV$ and $I_{\rm rel} < 0.15$~(0.2) are rejected.

For the lepton+jets and dilepton analyses, jets are reconstructed by clustering the PF candidates
using the anti-\kt algorithm~\cite{Cacciari:2008} implemented by
FastJet 3.0.4~\cite{Cacciari:2011ma} with a distance
parameter of $0.5$.
Charged PF particles that are inconsistent with the primary vertex with the highest
value of $\sum p^{2}_\mathrm{T}$ are removed from the clustering.
This requirement significantly suppresses contamination from charged
particles associated with pileup vertices.
The neutral component from pileup is removed by applying an estimated residual energy
correction based on the jet area~\cite{Cacciari:2007fd}.

Jets from $\PQb$ quark decays (b jets) are identified
with the
combined secondary vertex (CSV) b tagging algorithm~\cite{btvPerform}.
This is based on the presence of a displaced secondary
vertex in a jet, reconstructed from charged tracks, combined with other
quantities comprising track impact parameters, charged hadron kinematic variables,
track multiplicity, etc. The tight
CSV selection criteria (CSVT) with
a misidentification probability of 0.1\% for light-flavor jets with an
efficiency around 55\% for $\PQb$ jets is used.

The negative vector sum of the \pt of all the PF candidates ($\VEtmiss$) is calculated
for each event. The magnitude of $\VEtmiss$ ($\MET$)~\cite{CMS-PAS-PFT-09-001} is used in the lepton+jets
and dilepton analysis.

The all-hadronic channel uses a trigger that requires the scalar sum of the
transverse momenta of all jet candidates in the event ($\HT$) to be at least 750\GeV.
The lepton+jets channel uses a single-electron trigger with a
$\pt$ threshold of 27\GeV and
single-muon trigger with a $\pt$ threshold of 24\GeV.
The dilepton channel uses the dilepton ($\Pe \Pe$, $\Pe \Pgm$, and $\Pgm \Pgm$)
triggers, with leading and sub-leading lepton $\pt$ thresholds
of 17 and 8\GeV, respectively.

In the all-hadronic channel, the selected $\PW$ bosons and top quarks are
sufficiently energetic for their decay products to have a large Lorentz boost and
are reconstructed as single jets. Such jets are identified within $|\eta|<2.4$ using the jet
decomposition into subjets, followed by application of criteria based
on the kinematic properties of subjets. The Cambridge--Aachen
(CA) algorithm~\cite{CAcambridge}
with distance parameter of
$0.8$ is used to cluster jets that are considered for the $\PW$ boson
and top quark selections, instead of the anti-\kt algorithm that is used in the lepton+jets and dilepton analyses.

The identification of a boosted $\PW$ boson ($\PW$ tagging) attempts to identify
the two daughter quarks of the $\PW$ boson by using the N-subjettiness~\cite{Thaler:2011gf}
variable:
\begin{eqnarray}
	\tau_{\mathrm{N}} = \frac{1}{d_0}\sum_{i} {\pt}_ i\ \text{min}\{\Delta R_{1i},\Delta R_{2i},...,\Delta R_{Ni}\},
\end{eqnarray}
where $\Delta R_{ji}$ is the angular separation between the axis of the subjet
candidate $j$ and the axis of the constituent particle $i$,
and $d_{0}$ is a normalization
factor. The variable $\tau_{\mathrm N}$ is a
$\pt$-weighted angular distance from a jet constituent to the nearest subjet axis, and
is close to zero if a given jet is consistent with having N or fewer subjets.
The $\tau_{2}/\tau_{1}$ ratio is used to discriminate
between the signal $\PW$-tagged jets with two subjets and jets from light quarks and gluons with a
single hard subjet ($\tau_{2}/\tau_{1} < 0.5$). In addition, jet pruning~\cite{Ellis:2009su} is
used to remove soft and wide-angle radiation, which significantly
reduces the measured mass of QCD multijet events, while leaving the measured mass of $\PW$-tagged jets close
to the nominal $\PW$ boson mass.  The mass of the pruned jet is required to be
consistent with the $\PW$ boson mass ($ 70 < m_{\text{jet}} < 100\GeV$).
The difference in $\PW$ tagging efficiency between data and simulation
is corrected by a simulation-to-data scale factor derived from the
$\PW$+jets and dijet control samples~\cite{Khachatryan:2014vla}.

Boosted top quark identification ($\PQt$ tagging) discriminates signal
from background events by using the three-prong substructure of a
merged $\PQt$ jet. We use the CMS $\PQt$ tagging algorithm~\cite{Kaplan:2008ie}, which reclusters the jet until it finds
one to four subjets that
are consistent with daughters of the top quark decay~\cite{catop_cms}.
We require at least three subjets and determine the lowest mass $m_{\mathrm{min}}$
of the pairwise combinations of the three highest-\pt subjets.
This $m_{\mathrm{min}}$ is required to be
compatible with the mass of the $\PW$ boson ($m_{\mathrm{min}} > 50\GeV$).
Finally, the mass of the CA jet from the $\PQt$ tagging algorithm is required to be consistent with the top quark mass ($140 < m_{\text{jet}} < 250\GeV$).

The $\PQt$ tagging selection in this analysis also uses N-subjettiness discrimination.
In this case the variable of interest is $\tau_{3}/\tau_{2}$,
since t jets are expected to have three subjets ($\tau_{3}/\tau_{2}$ $<$ 0.55).
Exactly one of the three subjets originating from the top quark decay
should be a b jet,
which we identify by requiring the largest subjet CSV discriminator value
to satisfy the medium selection criteria.  This requirement has a misidentification probability of 0.1\% for light-flavor
jets and an efficiency of around 65\% for $\PQb$ jets~\cite{CMS-PAS-BTV-13-001}.

This $\PQt$ tagging algorithm was studied in lepton+jets data and
simulated samples enriched in top quarks with high Lorentz boost.
We use a simulation-to-data scale factor for $\PQt$ tagging derived
from these studies to correct the Monte Carlo (MC) samples in the all-hadronic channel ~\cite{JME13007}.

\section{Event selection and background estimation}

We search for the presence of a $\bs$ quark decaying to $\PQt \PW$ by looking for deviations
from the expected background in the distributions of kinematic variables
for the all-hadronic, lepton+jets, and dilepton channels.
The event selections and background estimations of these three channels are
presented below.

\subsection{All-hadronic channel}
The all-hadronic channel is characterized by a top quark and a $\PW$ boson, both of which
decay hadronically. After the trigger selection, exactly
two CA jets with $\pt$ of at least 425\GeV are required to be present
in the event. One high-$\pt$ jet is required to be W-tagged, while the
other is required to be $\PQt$-tagged.
The main backgrounds for this channel
are $\ttbar$ and multijet events, which are estimated using control regions in data.
The small background contribution
from single top quark production is estimated from simulation.

\begin{figure}[htb]
\begin{center}
\includegraphics[width=0.66\textwidth]{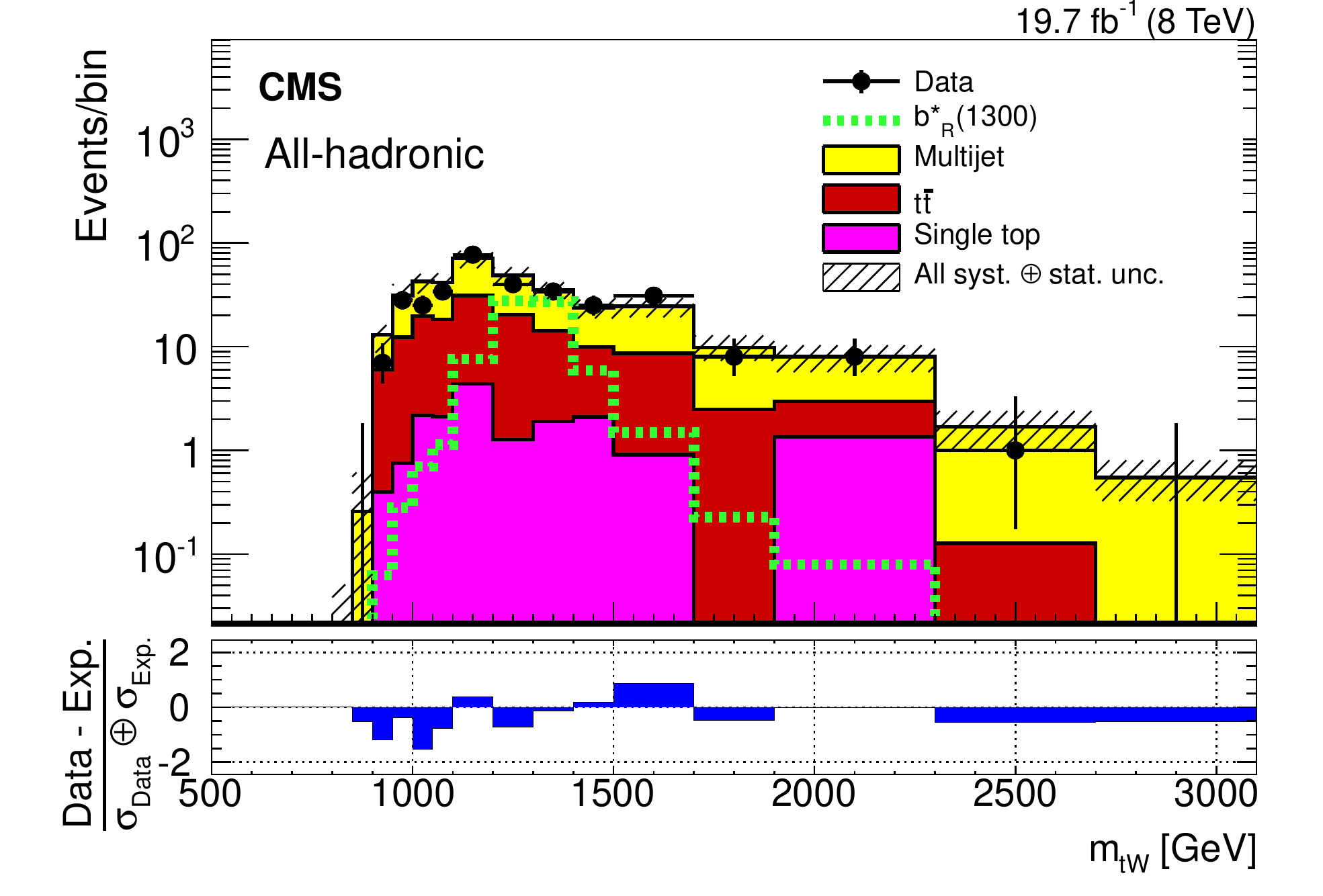}
\caption{
The invariant mass of the $\PQt \PW$ system in the all-hadronic channel after the full selection of data, the estimated background,
and the simulated signal with a $\bs$ mass of 1300\GeV. The combined statistical and
systematic uncertainties are indicated by the hatched band.  The bottom plot shows the pull ((data-background)/$\sigma_{\text{Data}}\oplus\sigma_{\text{Exp.}}$) between the data and the background estimate distributions.
The quantities $\sigma_{\text{Data}}$ and $\sigma_{\text{Exp.}}$ refer to the statistical uncertainty in data, and the systematic uncertainty in the background respectively.}
\label{figs:MtwvsBkg1}
\end{center}
\end{figure}

The multijet contribution is estimated by applying the top quark
mistagging ($\PQt$ mistagging) rate on events before $\PQt$ tagging is applied.
We measure the $\PQt$ mistagging rate using a control region where the contribution of signal events is
suppressed. For this control region we select
a $\PW$-tagged jet in the region of $30 < m_{\text{jet}} < 70\GeV$ or $m_{\text{jet}} > 100\GeV$.
After applying this selection we take the ratio of the number of jets
that are $\PQt$-tagged to the number of all top quark candidate jets
to define the $\PQt$ mistagging rate.
Here we use the $\PQt$ tagging algorithm
described in Section~\ref{objSel} but exclude the top quark candidate mass requirement that is applied to the
pre-tagged top quark candidate jets. The $\ttbar$ contamination is determined from
simulation and is accounted for when extracting the t mistagging rate.
The $\ttbar$ fraction in this region is about 25\% of the post-tag sample (numerator)
and 1\% of the pre-tag sample (denominator).
To extract a multijet background estimate, we weight the events that
pass the pre-$\PQt$-tagged selection
by the $\PQt$ mistagging rate. The parameterization of the $\PQt$ mistagging rate
is done as a function of the candidate jet $\pt$ and $\abs{\eta}$, in
order to account for kinematic correlations inherent in
$\PQt$ tagging. The mass distribution of the top quark candidate in the multijet
background estimate is corrected on a bin-by-bin basis by a weight extracted from
simulation, to correct for differences in the top quark candidate mass spectrum before and after $\PQt$ tagging.
This correction is such that it only changes the shape of the distribution
and has no effect on the overall normalization.
The correction factor depends on the mass of the top quark candidate and ranges from 0.45 at
low mass to 2.25 at high mass. The corresponding change in shape of the $m_{ \PQt \PW}$
spectrum is taken into account in the systematic uncertainties, and
makes a contribution that is much smaller than the total
systematic uncertainty, shown in Fig.~\ref{figs:MtwvsBkg1} as the hatched band.

The contribution from $\ttbar$ production is estimated by using a control region defined
by requiring one of the jets to pass inverted $\PW$ tagging requirements: $m_{\text{jet}}>130\GeV$ and
$\tau_{2}/\tau_{1} > 0.5$.
This selection has an enhanced $\ttbar$ fraction.
We compare the multijet and simulation-based $\ttbar$ background estimates to
the selection in data, then perform a fit to the invariant mass of the top quark candidate jet.
The template-based fit
constrains the multijet background template to move within its uncertainties, whereas the
normalization on $\ttbar$ is unconstrained. This study suggests that,
in addition to the scale factors that are applied, the $\ttbar$
contribution needs to be further scaled by $0.79 \pm 0.17$. The uncertainty in
this normalization is obtained from the fitting procedure.

The invariant mass of the top quark and $\PW$ boson candidate jets, $m_{ \PQt \PW}$, for the selected events in
the signal region is shown in Fig.~\ref{figs:MtwvsBkg1} and is used
for limit setting. The
expected number of events is $359 \pm 57$, and the observed number of events is 318 (Table~\ref{table:Cutflow}).

\begin{table}[htp]
\begin{center}
 \caption{Event yields in the all-hadronic channel after the final
   selection, normalized to an integrated luminosity of 19.7\fbinv.
Both statistical and systematic uncertainties are shown.
The systematic uncertainties are described in Section~\ref{sec:systematics}.
}
   \begin{tabular}{lr|x{10}}
  \hline
    \multicolumn{2}{l|}{Sample} & \multicolumn{1}{c}{Yield $\pm$ stat. $\pm$ syst.} \\
  \hline
$\bsL$ & 800\GeV  & 26.0   ,  1.9   \pm\, 7.4 \\
$\bsL$ & 1300\GeV & 57.8   ,  0.6   \pm\, 4.0 \\
$\bsL$ &1800\GeV & 4.1    ,  0.0   \pm\, 0.2 \\
$\bsR$ & 800\GeV  & 33.4   ,  2.2   \pm\, 9.1 \\
$\bsR$ & 1300\GeV & 72.5   ,  0.6   \pm\, 4.8 \\
$\bsR$ & 1800\GeV & 5.4    ,  0.0   \pm\, 0.3 \\
  \hline
    \multicolumn{2}{l|}{$\ttbar$}           & 129     ,  \ \ 3 \     \pm\, 42   \\
    \multicolumn{2}{l|}{Single top}         & 19.0    ,  2.9    \pm\, 6.5    \\
    \multicolumn{2}{l|}{Multijet }               & 211    , \ \ 0 \  \pm\, 38  \\
  \hline
  \multicolumn{2}{l|}{SM expected}       & 359    ,  \ \ 4 \  \pm\, 57  \\

  \hline
  \multicolumn{2}{l|}{Data} & \multicolumn{1}{c}{318}  \\
  \hline
   \end{tabular}
\label{table:Cutflow}
\end{center}
\end{table}

\subsection{Lepton+jets channel}
\label{sec:lepjets}

The lepton+jets channel
is characterized by the presence of
exactly one isolated electron or muon and a b jet, as well as at least two
light-flavor jets. The signal region is defined to have
exactly three jets with $\pt>40\GeV$ and $|\eta|<2.4$, together with
exactly one electron with $\pt>130\GeV$ and $|\eta|<2.4$, or exactly
one muon with $\pt>130\GeV$ and $|\eta|<2.1$.
Of these three jets, there must be exactly one jet that satisfies the CSVT b tagging
selection. The contributions of
the $\bs$ quark signal, $\ttbar$, single top quark, $\PZ$+jets,
and diboson processes are taken from simulation.
The multijet and $\PW$+jets background contributions are estimated from data.

The multijet background is estimated by performing a fit to the $\MET$
distribution for the electron channel, and a fit to the transverse
mass distribution of the leptonically decaying $\PW$ boson in the muon channel.
The choice of the variables used to estimate the background depends on the
accuracy with which they are modeled, the choice is different for the electron
and muon channels because different subdetectors are involved.
A multijet control sample is selected to model the multijet background
 distributions by reversing the lepton isolation selection criteria to
 $I_\text{rel}>0.3$; multijet events comprise $>$99\% of this sample. The
 other backgrounds are modeled using simulated events.
The multijet background from the control sample is normalized to the fitted yield
to model the multijet background distribution in the signal region.
The possibility of a small contamination
 from a signal is taken
into account in fitting the scale factors
to backgrounds involving $\PW$ bosons decaying leptonically.

The $\PW$+jets background is estimated by performing a template fit to the
distribution of the reconstructed invariant mass of the leptonically
decaying $\PW$ boson and a $\PQb$ jet, $m_{{\PQb} \ell \nu}$. The fit is performed separately
for the electron and muon channels.
The $p_x$ and $p_y$ of the neutrino from the $\PW$ boson decay are set equal to the $x$ and $y$ components
of the $\VEtmiss$. The $p_z$ component is estimated by
constraining the reconstructed mass of the $\PW$ boson to be 80.4\GeV~\cite{pdg},
resulting in two solutions.
If both solutions are real, the one with the lowest $|p_z|$ is selected.
If there is no real solution, $p_x$ and $p_y$ are varied until
there is a single solution that minimizes the distance between the neutrino
momentum and the missing momentum in the transverse plane. For the fit to the $m_{\PQb \ell \nu}$
distribution, the multijet background template is fixed to
the result of the multijet background estimated from data, with the shape taken from
the multijet-enriched control region. The SM $\ttbar$, single top quark, $\PZ$+jets,
and diboson templates are taken from the simulation with a common normalization
scale factor of $1.09\pm0.10$ obtained from the fit.
The $\PW$+jets template is taken from the simulation, and normalized to the fitted yield.
The possibility of a small contamination from a signal is taken into
account in the scale factors applied to
backgrounds with a top quark signature.

The expected $\bs$ quark signal and background events and observed data
events are listed in Table~\ref{eventYield} for the electron and
muon channels separately.

\begin{table}[htp]
\begin{center}
 \caption{Event yields in the lepton+jets channel after the final
   selection, normalized to an integrated luminosity of 19.7\fbinv.
Both statistical and systematic uncertainties are included.
The systematic uncertainties are described in Section~\ref{sec:systematics}.
}
   \begin{tabular}{lr|x{10}x{10}}

  \hline
   \multicolumn{2}{l|}{Sample} & \multicolumn{1}{c}{Yield $\pm$ stat. $\pm$ syst.} & \multicolumn{1}{l}{Yield $\pm$ stat. $\pm$ syst.} \\
  \multicolumn{2}{l|}{} & \multicolumn{1}{c}{Electron channel} & \multicolumn{1}{c}{Muon channel} \\
  \hline
    \bsL & 800\GeV  & 300   ,  \ \ 6 \    \pm\, 50  & 311   ,  \ \ 6 \    \pm\, 51 \\
    \bsL & 1300\GeV & 11.9  ,  0.2   \pm\, 3.3 & 12.7  ,  0.2   \pm\, 3.5\\
    \bsL & 1800\GeV & 0.8   ,  0.0   \pm\, 0.3 & 0.7   ,  0.0   \pm\, 0.3\\
    \bsR & 800\GeV  & 383   ,  \ \ 6 \     \pm\, 63  & 396   ,  \ \ 7 \    \pm\, 66 \\
    \bsR & 1300\GeV & 18.5  ,  0.2   \pm\, 5.0 & 18.2  ,  0.2   \pm\, 4.9\\
    \bsR & 1800\GeV & 1.0   ,  0.0   \pm\, 0.4 & 1.0   ,  0.0   \pm\, 0.4\\
  \hline
    \multicolumn{2}{l|}{$\ttbar$ }      & 2581   ,  \ 23 \   \pm\, 370 & 2736  , \ 23 \  \pm\, 400 \\
    \multicolumn{2}{l|}{Single top}     & 364    ,  \ \ \ 4 \   \pm\, 78  & 387   , \ \ \ 4 \   \pm\, 84 \\
    \multicolumn{2}{l|}{$\PW \PW$/$\PW \PZ$/$\PZ \PZ$}       & 17.9   , \ 1.2  \pm\, 2.7 & 19.4  , \ 1.4  \pm\, 3.4 \\
    \multicolumn{2}{l|}{$\PW$+jets}         & 671    ,  100  \pm\, 230 & 639   ,  \ 87 \  \pm\, 150 \\
    \multicolumn{2}{l|}{$\PZ$+jets}         & 92     ,  \ \ 15   \pm\, 33  & 80    ,  \ 13 \  \pm\, 33  \\
    \multicolumn{2}{l|}{Multijet}            & 678    ,  100  \pm\, 150 & \multicolumn{1}{c}{ $ \ \ 48 \ {}^{\ + \ 78}_{\ - \  48}  \ \  \pm\, 23 $ } \\
  \hline
  \multicolumn{2}{l|}{SM expected}   & 4404   ,  150  \pm\, 470 & 3909  ,  120   \pm\, 440 \\
  \hline
  \multicolumn{2}{l|}{Data} & \multicolumn{1}{c}{4368} & \multicolumn{1}{c}{3887} \\
  \hline
   \end{tabular}
\label{eventYield}
\end{center}

\end{table}

\begin{figure}[tb]
  \begin{center}
    \includegraphics[width=0.49\textwidth]{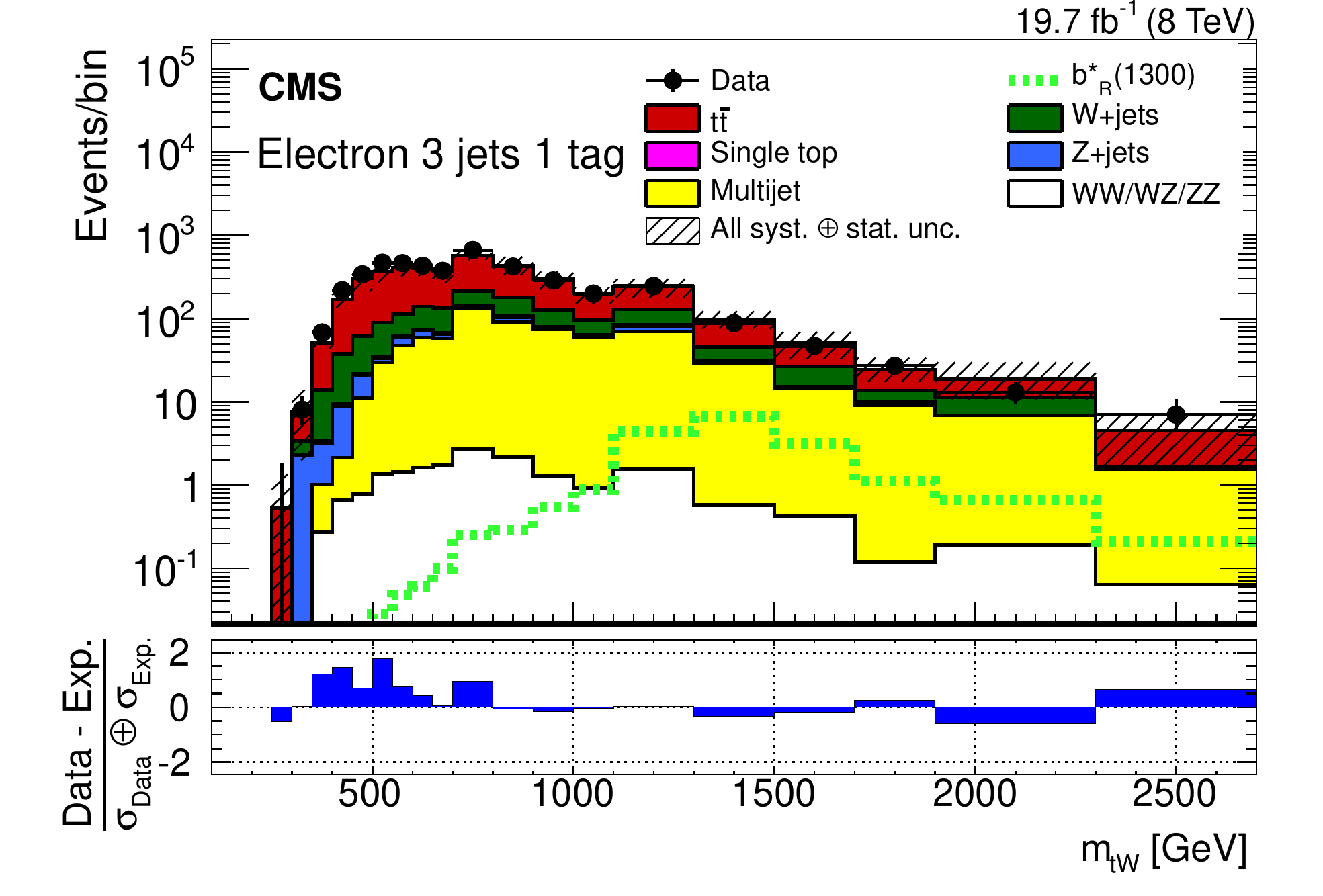}
    \includegraphics[width=0.49\textwidth]{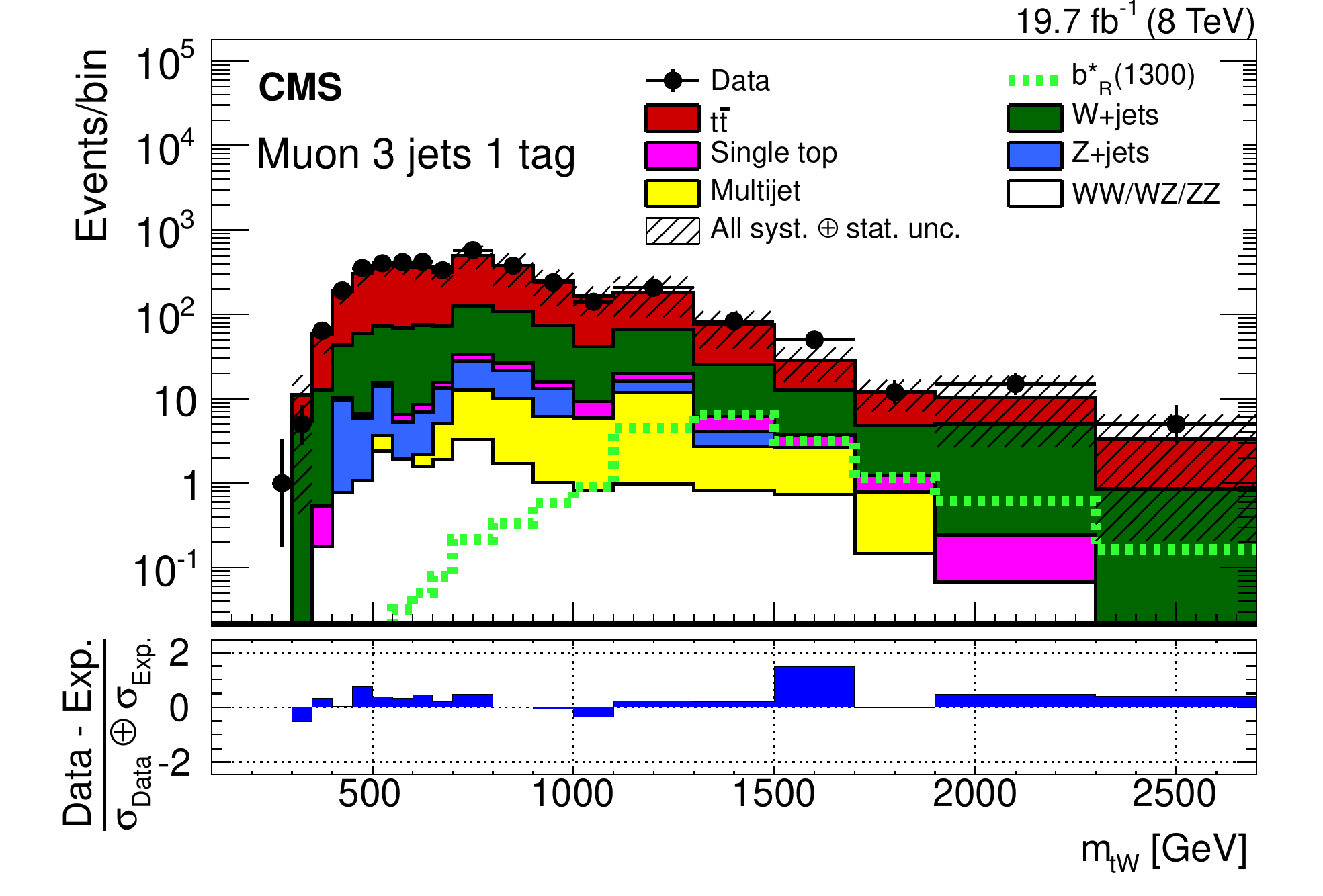}
 \caption{The invariant mass, $m_{\PQt \PW}$, in data compared to the SM background
  estimation for the electron (left) and muon (right) channels. The
  combined statistical and systematic uncertainties are indicated by the hatched band.  The bottom plots show the pull ((data-background)/$\sigma_{\text{Data}}\oplus\sigma_{\text{Exp.}}$) between the data and the background estimate distributions.
The quantities $\sigma_{\text{Data}}$ and $\sigma_{\text{Exp.}}$ refer to the statistical uncertainty in data, and the systematic uncertainty in the background, respectively.}
    \label{bstarMass}
  \end{center}
\end{figure}

We search for the $\bs$ signal as an excess above the predicted
backgrounds in the distribution of the invariant mass $m_{\PQt \PW}$
of the lepton, three jets, and $\VEtmiss$. In this calculation,
the neutrino $p_x$ and $p_y$ components are obtained from
$\VEtmiss$, and $p_z$ is set to zero since it cannot be measured by the detector
and could have multiple solutions from the analytical second order $W$ mass constraint.
The distribution of $m_{\PQt \PW}$ is shown in
Fig.~\ref{bstarMass}. The widths of the bins are chosen to
be comparable to the resolution in the reconstructed $m_{\PQt \PW}$.

\subsection{Dilepton channel}

The dilepton channel is characterized by two isolated,
oppositely charged electrons or muons and at least one jet. The signal
region is defined to have
at least one jet with $\pt > 30\GeV$ and $\abs{\eta} < 2.5$,
together with at least two leptons having $\pt > 30\GeV$
and $\abs{ \eta } < 2.5$ (2.4) for electrons (muons).
A minimum distance requirement of 0.3 between the two leptons in
$\Delta R$ removes photons radiated from muons in $\PW$+jets events, which can mimic
extra electrons. Most of the diboson background is removed by
requiring that the invariant mass of the two leptons is greater
than 120\GeV. In addition to the basic selections, events are required
to have $\MET > 40\GeV$. This requirement reduces
top quark background by 30\%, $\PW$+jets background by 50\%, diboson events
by 60\%, and removes over 95\% of $\PZ$+jets events, while keeping 90\%
of the signal events. The dominant backgrounds for
this channel are $\ttbar$, single top quark, $\PW$+jets, $\PZ$+jets, and diboson,
and are predicted by simulation.

A study is conducted to check the $\PW$+jets and multijet backgrounds using same-sign
events; the multijet background is found to be negligible, and the $\PW$+jets estimate agrees
with the MC simulation prediction within the statistical uncertainties.
Control regions, defined by reversing the \MET cut or by adding a $\PQb$
tagging requirement, are compared with data to confirm that
the dominant background sources are simulated correctly.

\begin{figure}[!h]
  \begin{center}
    \includegraphics[width=0.49\textwidth]{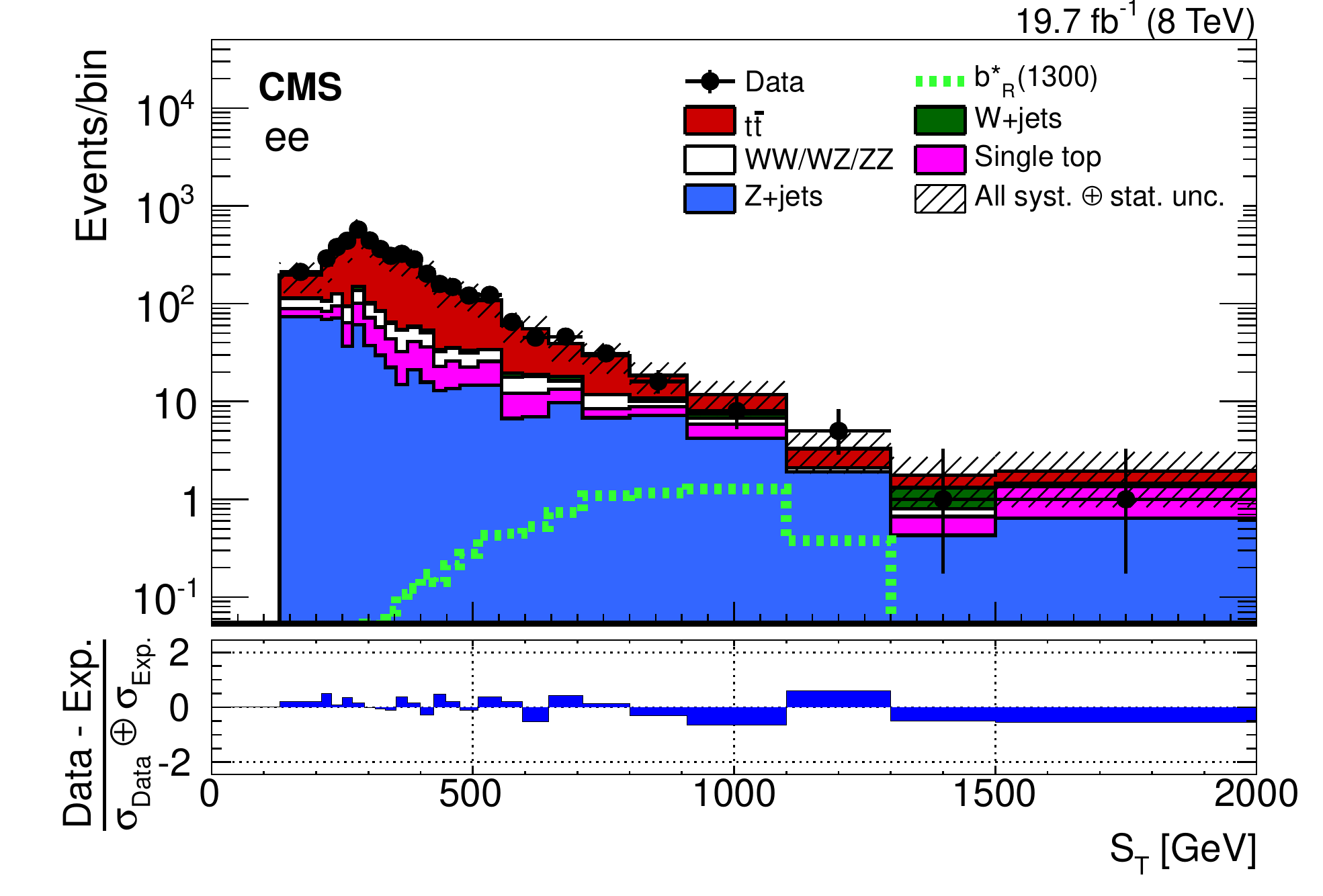}
    \includegraphics[width=0.49\textwidth]{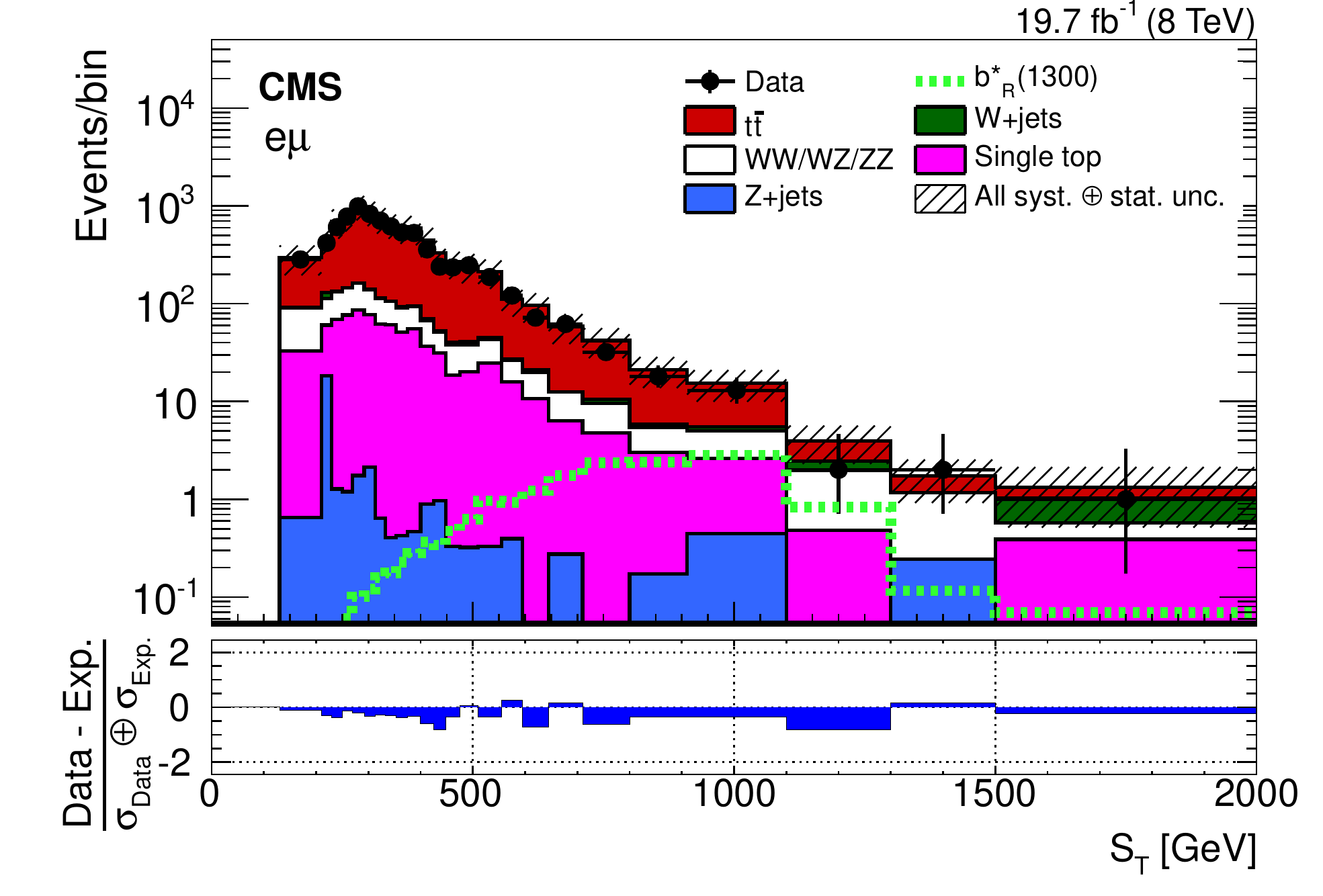}
    \includegraphics[width=0.49\textwidth]{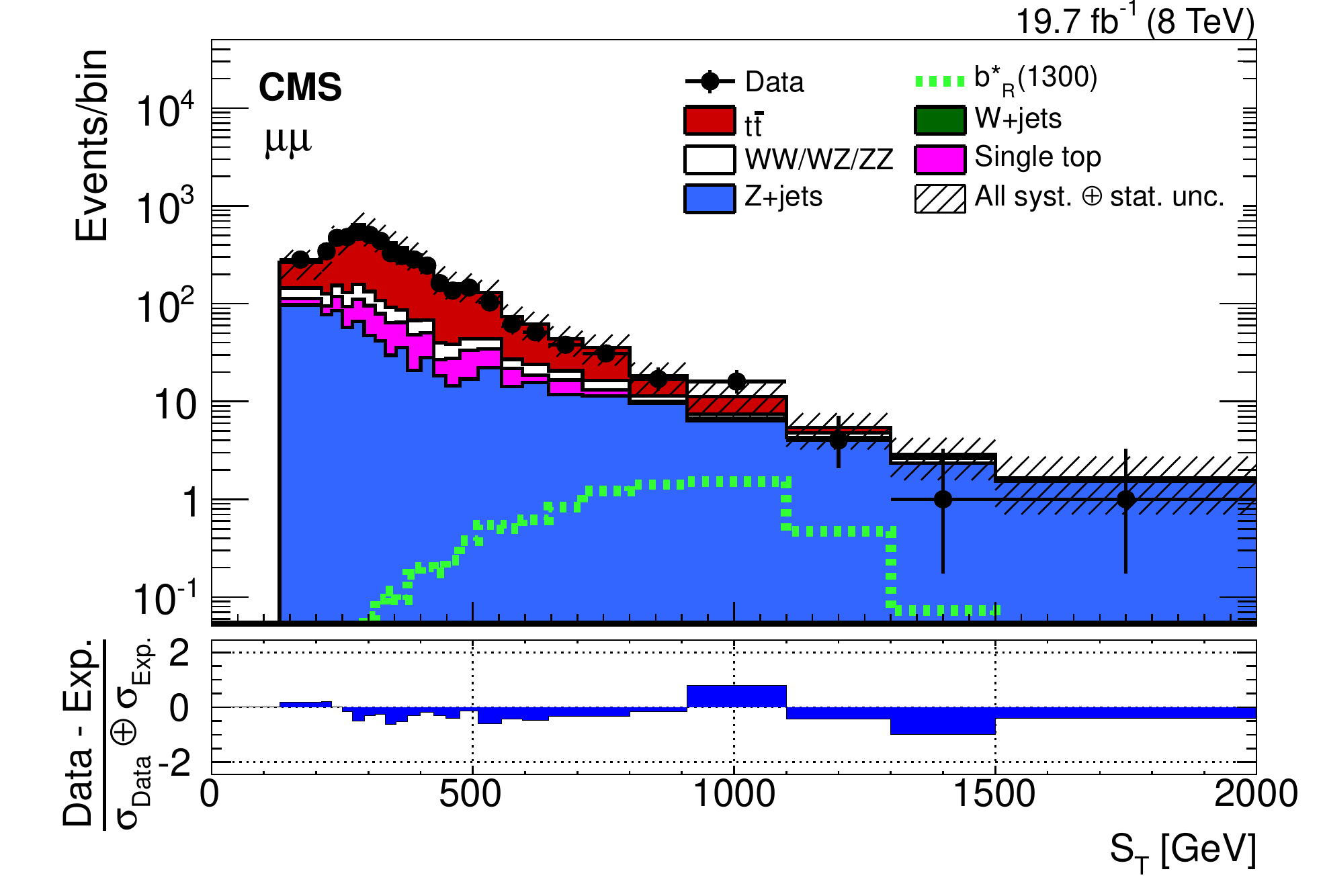}
    \includegraphics[width=0.49\textwidth]{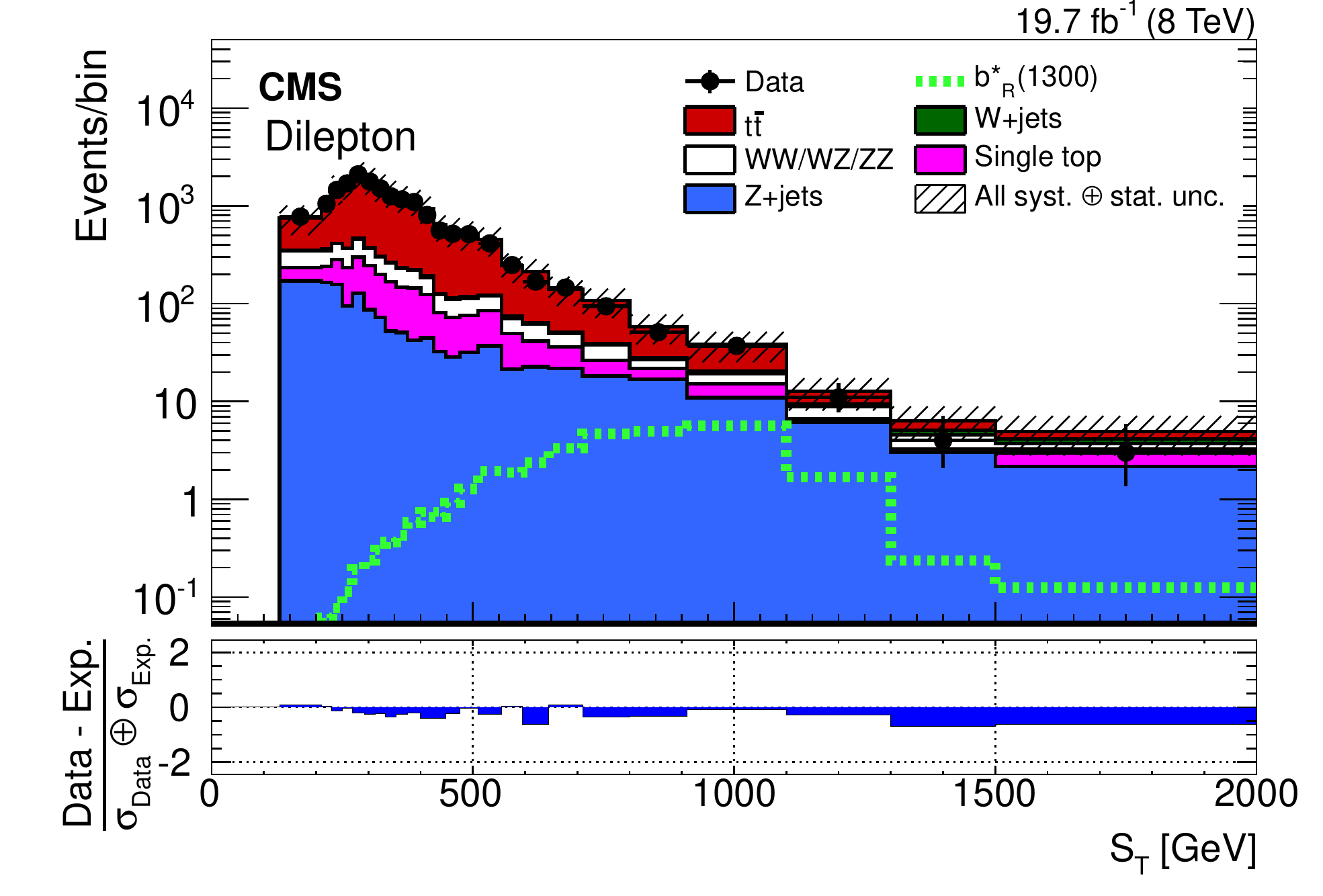}
   \caption{The $S_\mathrm{T}$ distribution
     for data and simulated samples after the event selection is
     applied, for $\Pe \Pe$ (top left), $\Pe \Pgm$  (top
     right), $ \Pgm \Pgm $ (bottom left), and inclusive dilepton (bottom right) channels.
     The combined statistical and
     systematic uncertainties are indicated by the hatched band.  The bottom plots show the pull ((data-background)/$\sigma_{\text{Data}}\oplus\sigma_{\text{Exp.}}$) between the data and the background estimate distributions.
     The symbols $\sigma_{\text{Data}}$ and $\sigma_{\text{Exp.}}$ refer to the statistical uncertainty in data, and the systematic uncertainty in the background, respectively.}
    \label{fig:final}
  \end{center}
\end{figure}

We search for the $\bs$ quark signal events using the distribution of
the scalar sum $S_\mathrm{T}$ of the $\pt$ of the two leading leptons, the jet with
the highest $\pt$, and \MET.
The distribution of this variable is shown in Fig.~\ref{fig:final}.
The results of the full selection are listed in Table~\ref{tab:Selection}.

\begin{table}[htp]
\begin{center}
 \caption{Event yields for the dilepton channel after the final selection, normalized to
   an integrated luminosity of 19.7\fbinv. Both statistical and systematic
   uncertainties are included. The systematic uncertainties are described in Section~\ref{sec:systematics}.}
   \begin{tabular}{lr|x{10}x{10}x{10}}
  \hline
    \multicolumn{2}{l|}{Sample} & \multicolumn{1}{c}{Yield $\pm$ stat. $\pm$ syst. } &  \multicolumn{1}{c}{ Yield $\pm$ stat. $\pm$ syst. } &  \multicolumn{1}{c}{ Yield $\pm$ stat. $\pm$ syst. } \\
     \multicolumn{2}{l|}{}                    & \multicolumn{1}{c}{$\Pe \Pe$ channel } & \multicolumn{1}{c}{ $\Pe \Pgm$ channel } & \multicolumn{1}{c}{ $\Pgm \Pgm$ channel } \\ \hline
\bsL & 800\GeV  &	158  ,  \ \ 2 \ \pm \,	32  & 347   ,  \ \ 3 \   \pm \, 72  & 192  ,  \ \ 3 \   \pm \, 39 \\
\bsL & 1300\GeV &	6.4  ,  0.1 \pm \,	1.5 & 14.3  ,  0.1  \pm \, 3.3 & 7.7  ,  0.1  \pm \, 1.7\\
\bsL & 1800\GeV &	0.4  ,  0.0 \pm \,	0.1 & 0.8   ,  0.0  \pm \, 0.2 & 0.5  ,  0.0  \pm \, 0.1\\
\bsR & 800\GeV  &	203  ,  \ \ 2 \  \pm \,	42  & 452   ,  \ \ 4 \    \pm \, 94  & 243  ,  \ \ 3 \   \pm \, 50 \\
\bsR & 1300\GeV &	7.4  ,  0.1 \pm \,	1.7 & 16.5  ,  0.1  \pm \, 3.7 & 8.9  ,  0.1  \pm \, 2.0\\
\bsR & 1800\GeV &	0.4  ,  0.0 \pm \,	0.1 & 0.9   ,  0.0  \pm \, 0.2 & 0.5  ,  0.0  \pm \, 0.1\\
\hline
 \multicolumn{2}{l|}{  $\ttbar$  }     &	3157  ,  24   \pm \,	530&	7226  ,  40   \pm \, 1220 & 3939  , 	29   \pm \,	660\\
 \multicolumn{2}{l|}{  Single top }    &	323   ,  12   \pm \,	83&	775   ,  19   \pm \, 210  & 414   , 	14   \pm \,	110\\
 \multicolumn{2}{l|}{  $\PW \PW$/$\PW \PZ$/$\PZ \PZ$  }     &	323   ,  \ 5 \   \pm \,	110&	700   ,  \ 2 \   \pm \, 240  & 399   , 	10   \pm \,	130\\
 \multicolumn{2}{l|}{ $\PW$+jets  }       &	38  ,  12   \pm \,	3.2&	45  ,  15   \pm \, 1.4  & 1   , 	0.4  \pm \,	0.0\\
 \multicolumn{2}{l|}{ $\PZ$+jets   }       &	553   ,  24   \pm \,	130&	31.6  ,  5.0  \pm \, 5.4  & 734   , 	29   \pm \,	170\\
 \hline
  \multicolumn{2}{l|}{SM expected }   &	4396  ,  38  \pm \,	558&	8777  , 	47   \pm \,	1257  & 5487  , 	45  \pm \,	699\\
 \hline
 \multicolumn{2}{l|}{Data 	}	&	 \multicolumn{1}{c}{ 4583	}	&	\multicolumn{1}{c}{7873}	& \multicolumn{1}{c}{4988} \\
  \hline
   \end{tabular}
\label{tab:Selection}
\end{center}
\end{table}

\section{Systematic uncertainties}
\label{sec:systematics}
Systematic uncertainties are divided into four groups:
theoretical, background normalization, instrumental, and other
measurement-related uncertainties. These uncertainties are summarized
in Table~\ref{tab:corr_systematics}.

\subsection{Theoretical uncertainties}

Several uncertainties in event simulation are considered. The PDF uncertainties are estimated with the CT10 PDF
  eigenvector set~\cite{PDF:CTEQ10}.

  In order to estimate uncertainties originating from the top quark mass,
  additional simulated samples are produced by varying the top quark mass up
  and down by 5\GeV. A linear extrapolation is applied to scale down the top quark mass
  uncertainty to 1\GeV. This is applied to $\ttbar$ and
  single top quark $t$-channel samples.
In order to estimate uncertainties originating from the
choice of the renormalization and factorization
  scales ($\mu_{R}$ and $\mu_{F}$), for the $\ttbar$, and single top quark $t$-channel simulation, the nominal samples
  use
  $\mu_{R}^2$ = $\mu_{F}^2$ = $M^{2}_{\PQt}+\sum
  p_{\mathrm{T}}^{2}$~\cite{madgraph5},
  where $\sum p_{\mathrm{T}}^{2}$ sums over outgoing partons.
  To evaluate the effect of this scale choice, additional MC samples are produced by
  varying $\mu_{R}$ and $\mu_{F}$ simultaneously by a factor of 0.5 or 2.0.

\subsection{Background normalization uncertainty}

  For the lepton+jets and dilepton analysis, the $\ttbar$ cross section uncertainty of $\pm 5.3\%$~\cite{Chatrchyan:2013faa} is used.

  The all-hadronic channel extracts the $\ttbar$ normalization from data,
  resulting in an uncertainty of 22\% obtained from the fit.

The normalization uncertainties in single top quark $t$-, $\PQt \PW$-, and $s$-channel cross
sections are 15\%, 30\% and
20\%, respectively~\cite{Kidonakis:2012db}. The normalization
uncertainties in diboson production cross sections
   are 30\%, which is the sum of the experimentally measured cross section uncertainty~\cite{CMS-SMP-12-024}
   and uncertainties due to extra jet production. The normalization uncertainty in
   the $\PZ$+jets background is 20\%, which is the sum of the experimentally measured
   cross section uncertainty~\cite{Chatrchyan:2014mua} and uncertainties due to extra jet production. The
   normalization uncertainty in the $\PW$+jets background is 45\% for
   the electron+jets channel and 30\% for the muon+jets channel, estimated
   from data and described in Section~\ref{sec:lepjets}. Detector effects
   and modeling uncertainties that affect the templates are included in the uncertainty.

  The normalization uncertainties in the multijet backgrounds are $\pm 33\%$
  and $^{+170}_{-100}$\% for the electron+jets and the muon+jets
  channels, respectively, estimated from data and described in Section~\ref{sec:lepjets}.
The uncertainties originating from detector effects, theoretical modeling,
and the multijet background control
region choice are summed in quadrature to give the
uncertainties in the multijet and $\PW$+jets background estimations.

\subsection{Other measurement uncertainties}

  In the all-hadronic channel, we correct the simulation by using the trigger efficiency extracted from data that is
  obtained from a control sample triggered with a lower $\HT$ threshold than in the standard event selection.
  The scale factors are parameterized as a function of the summed leading and sub-leading jet $\pt$.
  To obtain a systematic uncertainty for this correction,
we vary the trigger efficiency $\epsilon$ by
$\pm (1-\epsilon)/2 $, which results in less
than a 1\% change of the yields for all samples.

  The differences between data and simulation due to the electron trigger, identification, and isolation efficiencies
 are corrected with $\pt$- and $\eta$-dependent scale factors by comparing simulation with a
  $\PZ \to \Pe \Pe$ data sample. The uncertainties due
  to the statistically limited $\PZ \to \Pe \Pe$
  samples and the uncertainties in the theoretical inputs to the
  simulation are taken into account.

The scale factor measurements define the uncertainties for electron trigger,
  identification, and isolation requirements, and these uncertainties are less than 1\%.
  Scale factors related to the muon trigger, identification, and
  isolation efficiency are measured in a similar way to those for electrons, but
  use $\PZ \to \Pgm \Pgm$, where the uncertainties are less than
  2\%~\cite{Chatrchyan:2012xi}, instead of $\PZ \to \Pe \Pe$ events.

The jet energy resolution~\cite{Chatrchyan:2011ds} systematic uncertainty is an
$\eta$-dependent smearing of the jet energy resolution for simulated
events, which results in a less than 0.4\% acceptance change. The jet energy scale
~\cite{Chatrchyan:2011ds} systematic uncertainty is parameterized in
$\pt$ and $\eta$
and applied to simulated samples to cover the difference between data
and simulation, which is typically 5\% or less. The all-hadronic channel has an
additional 3\% uncertainty because the jet energy scale is measured from anti-\kt jets,
but applied to CA jets. The jet energy scale uncertainty is propagated in the $\MET$ calculation.
The estimation of \MET includes an additional uncertainty due to the effect of unclustered energy
arising from the jets or leptons.

  The b tagging efficiency and mistagging rate uncertainty are estimated by comparing a b jet enriched
  $\Pgm$+jets data sample with simulation~\cite{CMS-PAS-BTV-13-001}.
  The differences are corrected by jet flavor ($\PQb$ jet, $\PQc$ jet, and light
  jets from $\PQu$/$\PQd$/$\PQs$/gluon), $\pt$- and $\eta$-dependent $\PQb$ tagging and mistagging scale
  factors. The uncertainties in these scale factors are propagated
  to the $\PQb$ tagging event weight calculation independently, giving
  the uncertainties in $\PQb$ tagging efficiency and mistagging rate. The typical acceptance
  change due to $\PQb$ tagging efficiency is less than 3\%. The mistagging rate brings
  about an uncertainty of 0.3\% for samples that have at least one $\PQb$ jet and of 9.0\% for
  samples that have no $\PQb$ jets. The all-hadronic channel includes a $13\%$ uncertainty in the $\PQt$ tagging
scale factor, which is used to correct for differences
in subjet identification efficiencies between data and simulation~\cite{JME13007}.

The result of a polynomial fit to the $\PQt$ mistagging rate extracted from
a control region as a function of jet $\pt$
is applied to events before applying the $\PQt$ tagging algorithm to estimate
the multijet background contribution in the all-hadronic channel. The
fit introduces a 9\% statistical uncertainty and a
12\% uncertainty to allow for the possibility of choosing alternative
functional forms.
There is a difference between the shape of the jet mass distribution of the top
quark candidate in the control and signal regions.
This is corrected by a top quark jet mass
dependent weight derived from the multijet simulation.  This correction contributes an
extra 0.3\% uncertainty in the total multijet yield. The uncertainty due to the
choice of parameterization in the $\PQt$ mistagging rate is taken to be the difference
between a parameterization in $\pt$, $\eta$ and a
parameterization in $\pt$, $\eta$, $m_{\PQt \PW}$. This difference is about
2\% of the total multijet yield, with an additional 20\% statistical uncertainty from
the higher dimensional parameterization.

To estimate the uncertainty due to pileup modeling in simulation, we
vary the measured minimum bias cross section of 69.4\unit{mb} by $\pm
5\%$. These variations are then propagated to analysis results by
modifying the pileup multiplicity
accordingly~\cite{Chatrchyan:2012nj}. The uncertainty in the
integrated luminosity is measured in dedicated samples and
applied to the signal and backgrounds based on simulation. The size of
this uncertainty is 2.6\%~\cite{lumi2013}.

\begin{table}
\begin{center}
\caption{Sources of systematic uncertainty for the three analysis channels. For the shape-based uncertainties,
 the parameterization used for the uncertainty deviation is given in parentheses.
 Sources marked with "sideband" are measured from data, and contain various uncertainty sources.
 Uncorrelated uncertainties that apply to a given channel are
 marked by $\odot$. Uncertainties correlated between channels
 are marked by $\oplus$. The uncertainties varying as functions of variables in question
 are indicated if no uncertainty value is listed.
}

 \label{tab:corr_systematics}
\begin{tabular}{l c | c c c c }
\hline
Source of uncertainty   & Uncertainty & All-   &  Lepton+ & Dilepton  \\
  & & hadronic   &  jets &  \\
\hline
Integrated luminosity                       & \multicolumn{1}{P|}{2.6p}                      & $\oplus$  & $\oplus$ & $\oplus$  \\
\ttbar cross section                        & \multicolumn{1}{P|}{5.3p}                      &           & $\oplus$ & $\oplus$  \\
\ttbar normalization from data              & \multicolumn{1}{P|}{22p}                       & $\odot$   &          &           \\
Single top quark $t$-channel $\sigma$       & \multicolumn{1}{P|}{15p}                       & $\oplus$  & $\oplus$ &  \\
Single top quark $\PQt \PW$-channel $\sigma$        & \multicolumn{1}{P|}{20p}                       & $\oplus$  & $\oplus$ & $\oplus$  \\
Single top quark $s$-channel $\sigma$       & \multicolumn{1}{P|}{30p}                       & $\oplus$  & $\oplus$ &  \\
Diboson cross section                       & \multicolumn{1}{P|}{30p}                       &           & $\oplus$ & $\oplus$  \\
$\Z$+jets cross section                     & \multicolumn{1}{P|}{20p}                      &           & $\oplus$ & $\oplus$  \\
$\PW$+jets cross section                    &  \multicolumn{1}{P|}{8p}                       &           &          & $\odot$  \\
Double lepton triggers                      &  \multicolumn{1}{P|}{2p}   &           &  & $\odot$   \\
Dilepton muon ID and isolation              &  \multicolumn{1}{P|}{2p}   &           &  & $\odot$   \\
Dilepton electron ID and isolation          &  \multicolumn{1}{P|}{2p}   &           &  & $\odot$   \\
Dilepton pileup uncertainty                 & \multicolumn{1}{P|}{2.6p}  &           &  & $\odot$   \\
$\PW$ tagging                                   & \multicolumn{1}{P|}{8p}                        & $\odot$   &          &           \\
$\PQt$ tagging                                   & \multicolumn{1}{P|}{13p}                       & $\odot$   &          &           \\
Unclustered energy ($\MET$ uncertainty)     & \multicolumn{1}{P|}{10p}                       &           & $\odot$ &  \\
\hline	
Single-lepton triggers                      & ${\pm}1 \sigma (\pt,\eta)$  &           & $\odot$  &           \\
\HT trigger                                 & \small{${\pm}1 \sigma (p_{\mathrm{T1}}+p_{\mathrm{T2}})$}  & $\odot$   &          &           \\
Electron ID and isolation        & ${\pm}1 \sigma (\pt,\eta)$   &           & $\odot$ &  \\
Muon ID and isolation            & ${\pm}1 \sigma (\pt,\eta)$   &           & $\odot$ &  \\
Jet energy scale                            & ${\pm}1 \sigma (\pt,\eta)$   & $\oplus$  & $\oplus$ & $\oplus$  \\
Jet energy resolution                       & ${\pm}1 \sigma (\eta)$       & $\oplus$  & $\oplus$ & $\oplus$  \\
Pileup uncertainty                          & ${\pm}1 \sigma$              &           & $\odot$ &  \\
$\PQb$ tagging efficiency                        & ${\pm}1 \sigma (\pt, \eta)$  &           & $\odot$  &           \\
$\PQb$ mistagging rate                           & ${\pm}1 \sigma (\pt,\eta)$   &           & $\odot$  &           \\
Multijet background                         & sideband                   & $\odot$   & $\odot$  &  \\
$\PW$+jets background                       & sideband                   &           & $\odot$  &           \\
\hline
PDF uncertainty                             & ${\pm}1 \sigma$              &   & $\odot$ &   \\
\ttbar $\mu_{R}$ and $\mu_{F}$ scales       & \small{4$Q^2$ and 0.25$Q^2$}       & $\oplus$  & $\oplus$ &  $\oplus$ \\
Top quark mass                           & \small{$\pm$ 1\GeV for $\rm{m_{top}}$}      &           & $\odot$  &           \\
\hline
Simulation statistical uncertainty                  &                            & $\odot$   & $\odot$  & $\odot$   \\
\hline

\end{tabular}
\end{center}
\end{table}

\section{Results and interpretation}

\begin{figure}[phtb]
\centering
\includegraphics[width=0.48\textwidth]{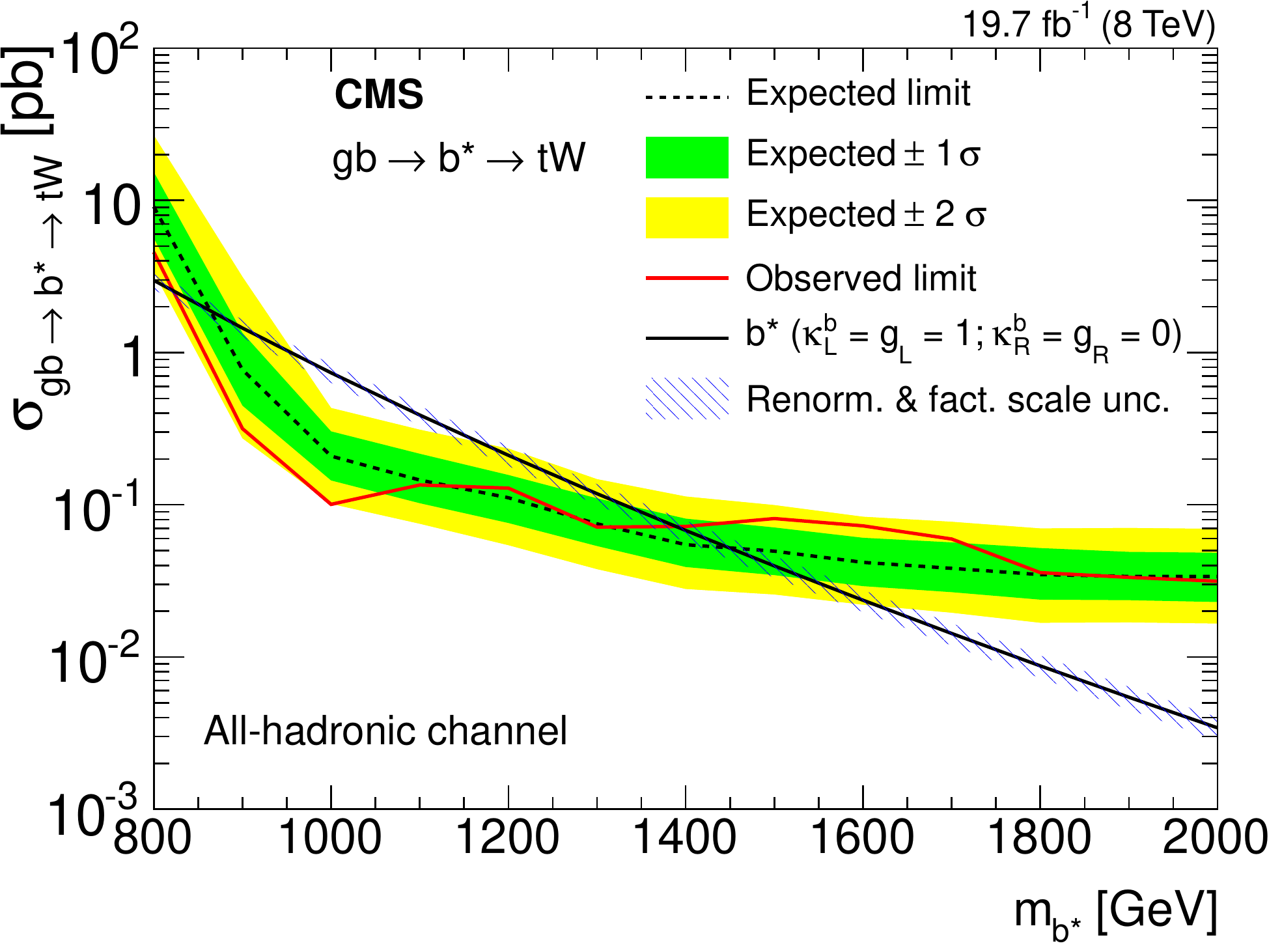}
\includegraphics[width=0.48\textwidth]{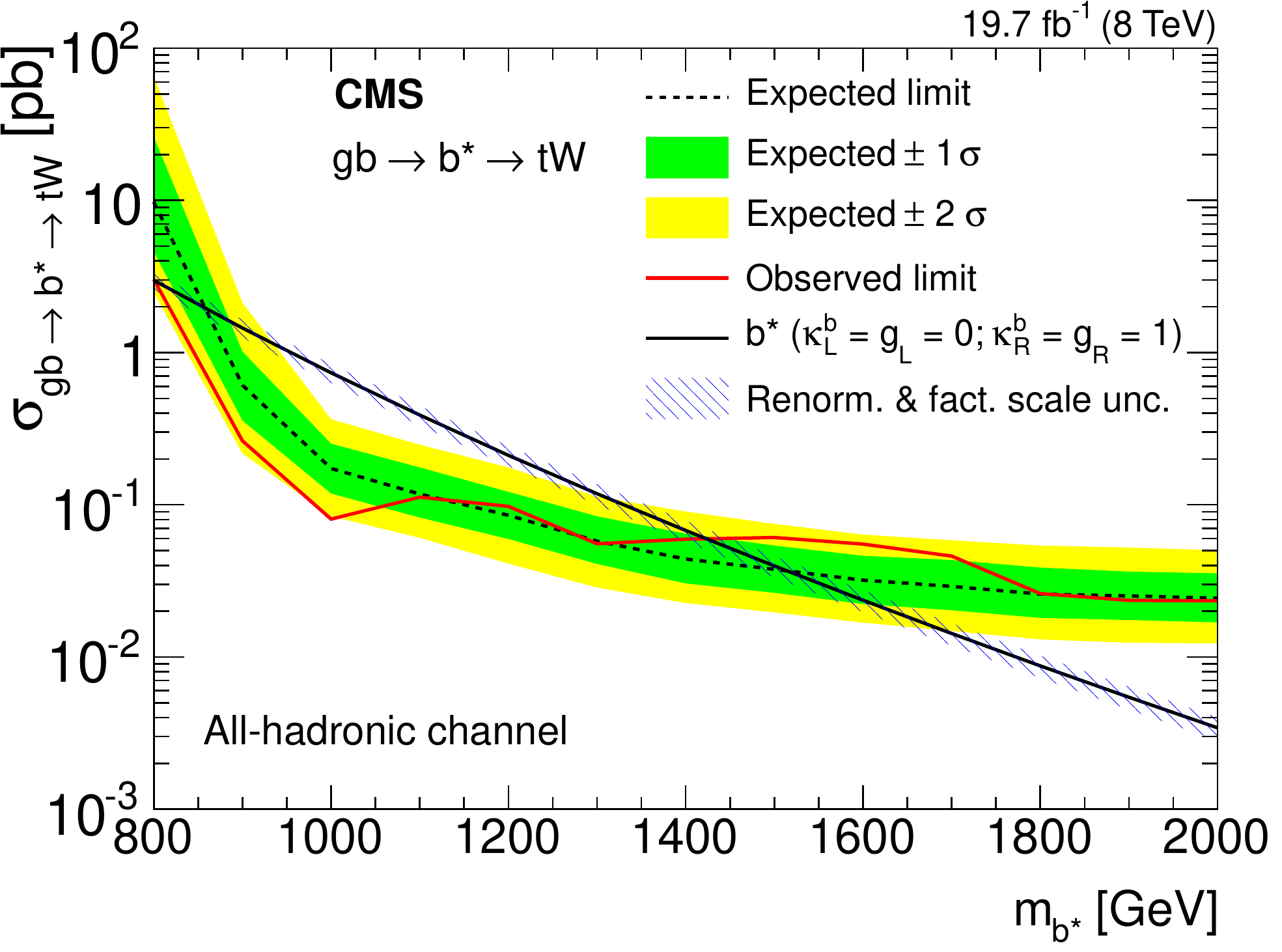}
\includegraphics[width=0.48\textwidth]{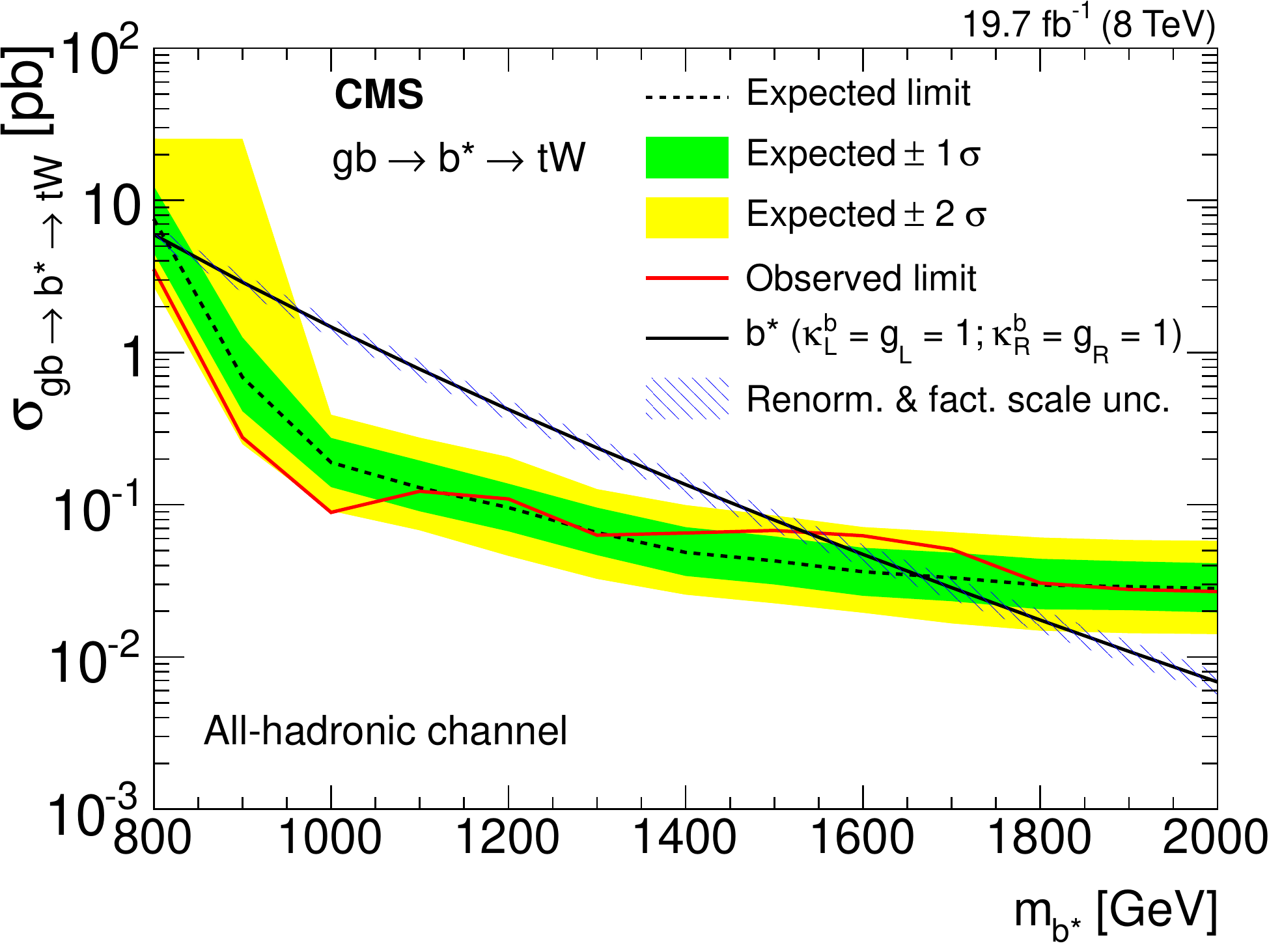}
\caption{The expected (dashed) and observed (solid) production cross section limits at 95\% CL for the all-hadronic channel as a function of $\bs$ quark mass for $ \Pg \PQb \to \bs \to \PQt \PW$. The theoretical cross section (solid line with hatched area) is also shown for comparison. The $1\sigma$ and $2\sigma$ uncertainties in the expected limit bands are shown. Limits for the left-handed, right-handed, and vector-like $\bs$ quark coupling hypotheses are shown in the top left, top right, and bottom plots, respectively.
\label{fullHadlimitPlot}}
 \end{figure}

\begin{figure}[phtb]
\centering
\includegraphics[width=0.48\textwidth]{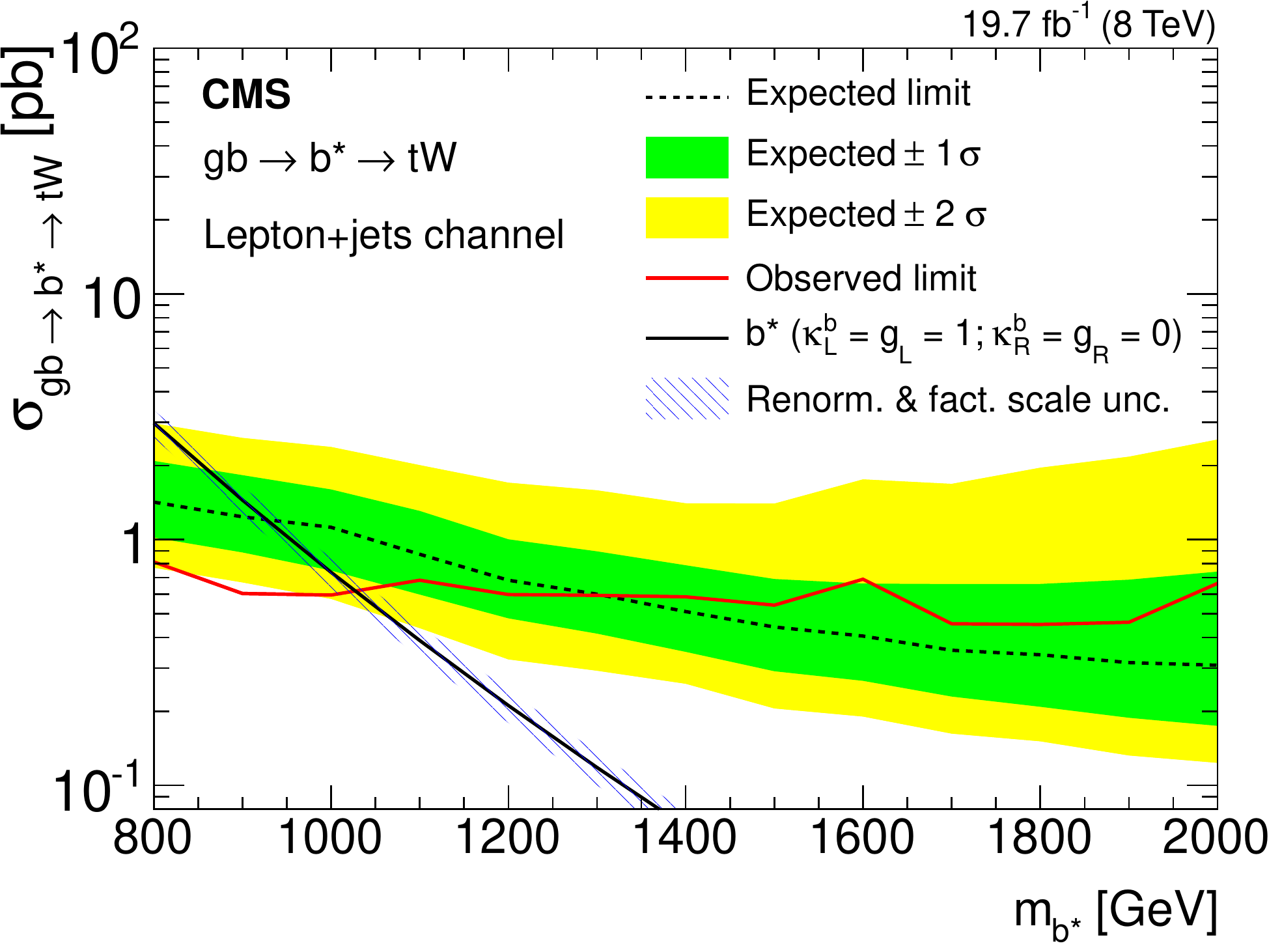}
\includegraphics[width=0.48\textwidth]{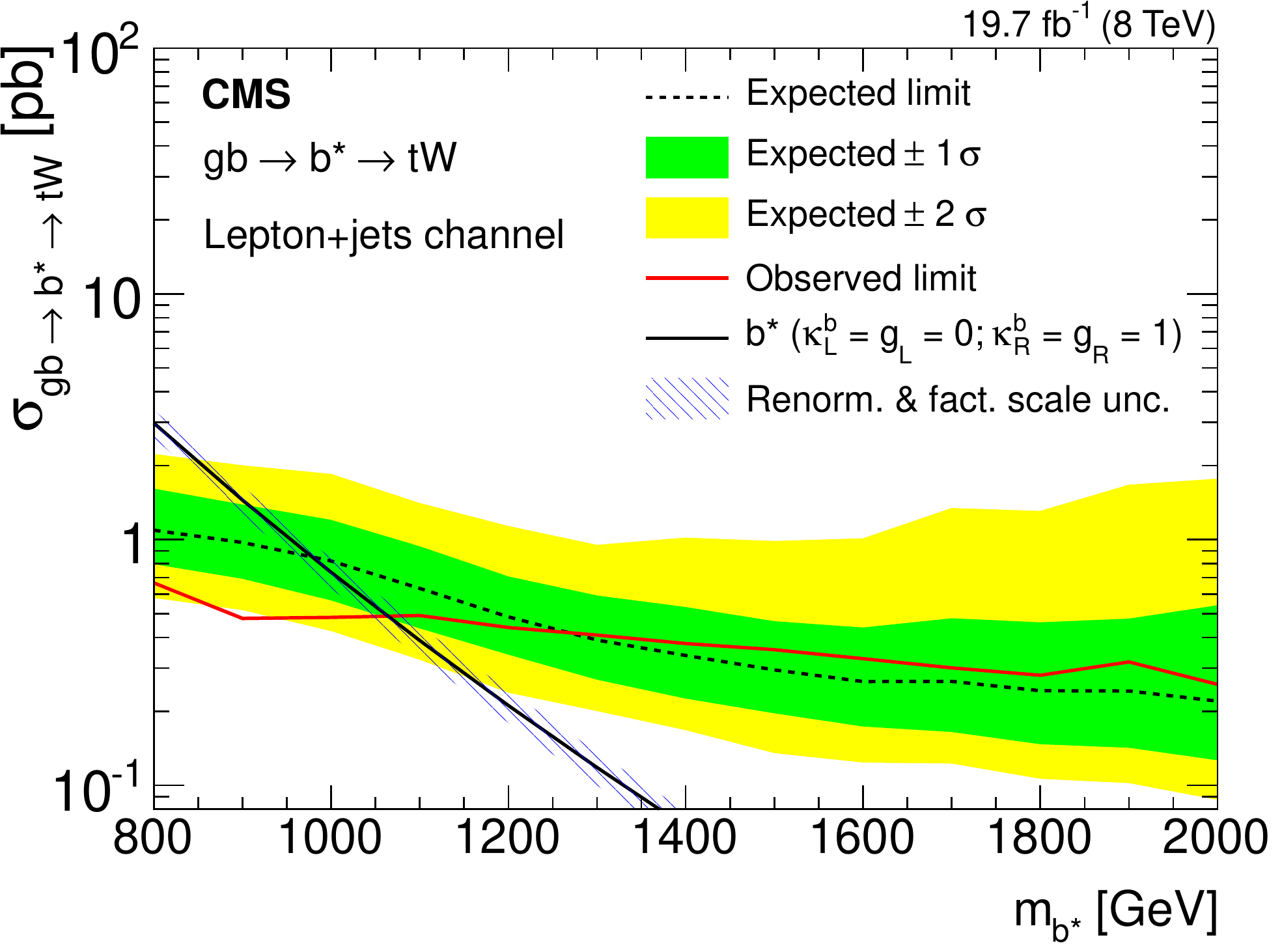}
\includegraphics[width=0.48\textwidth]{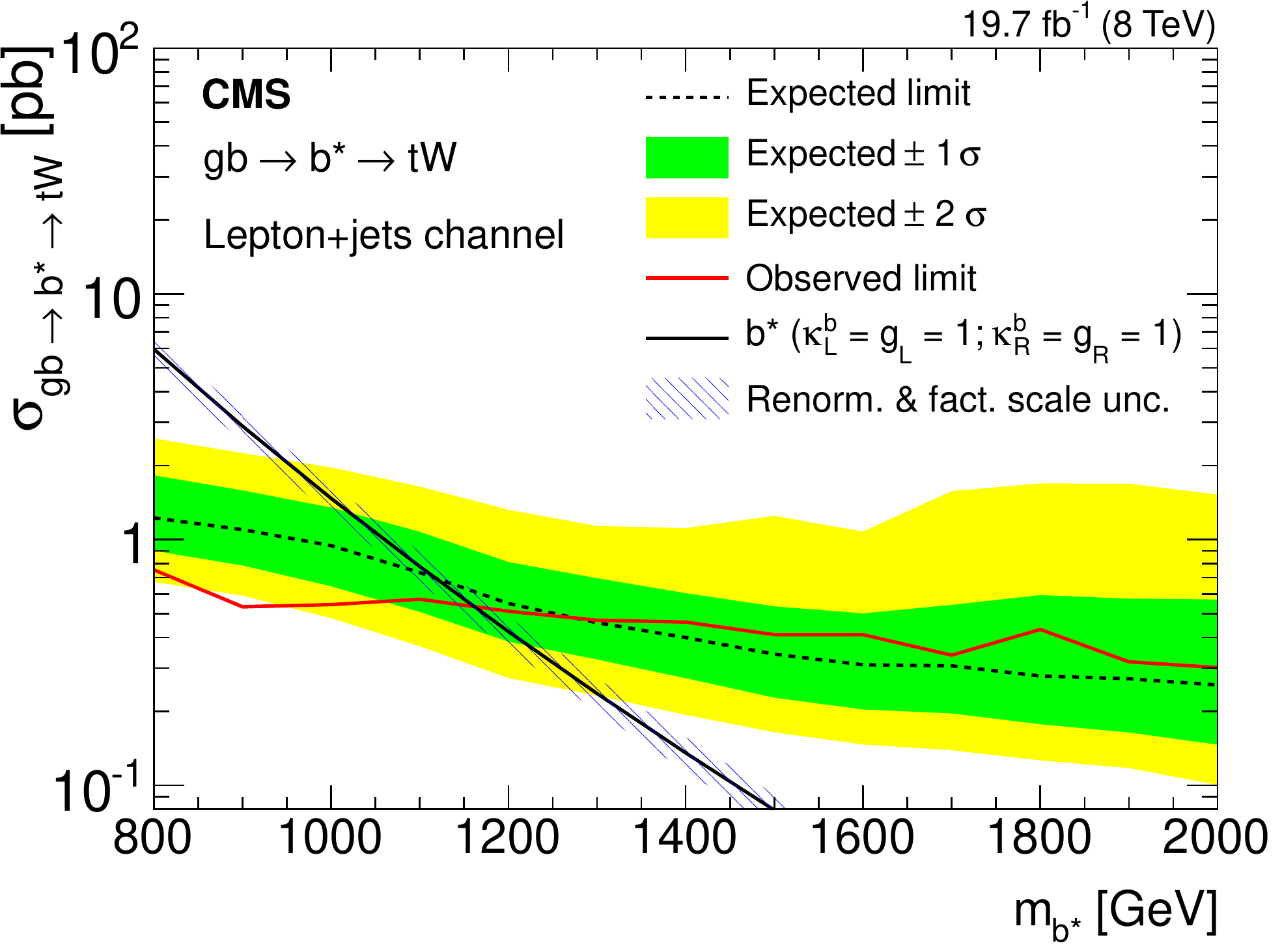}
\caption{The expected (dashed) and observed (solid) production cross section limits at 95\% CL for the lepton+jets channel as a function of $\bs$ quark mass for $ \Pg \PQb \to \bs \to \PQt \PW$. The theoretical cross section (solid line with hatched area) is also shown for comparison. The $1\sigma$ and $2\sigma$ uncertainties in the expected limit bands are shown. Limits for the left-handed, right-handed, and vector-like $\bs$ quark coupling hypotheses are shown in the top left, top right, and bottom plots, respectively.
\label{lepjetlimitPlot}}
\end{figure}

\begin{figure}[phtb]
\centering
\includegraphics[width=0.48\textwidth]{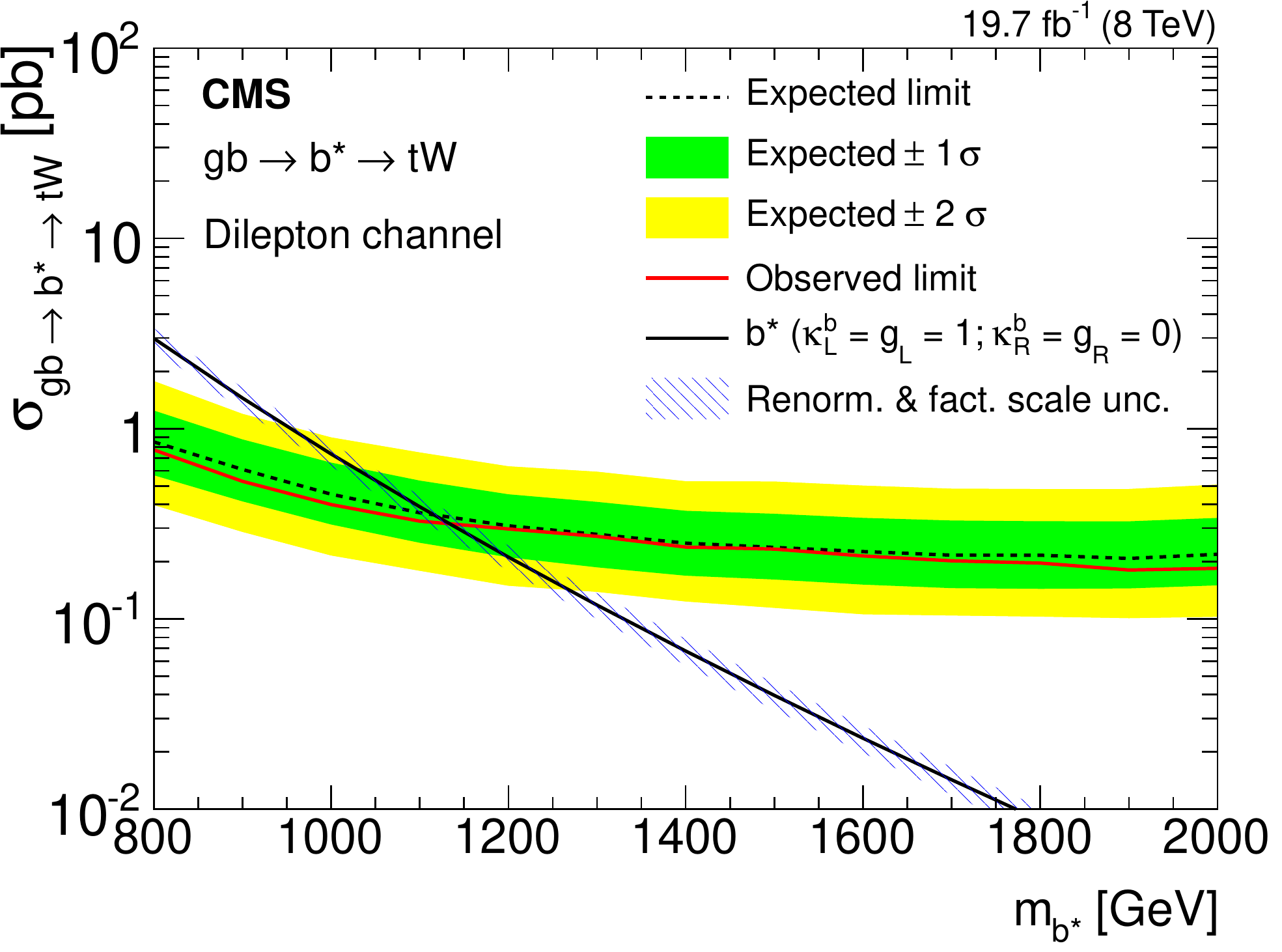}
\includegraphics[width=0.48\textwidth]{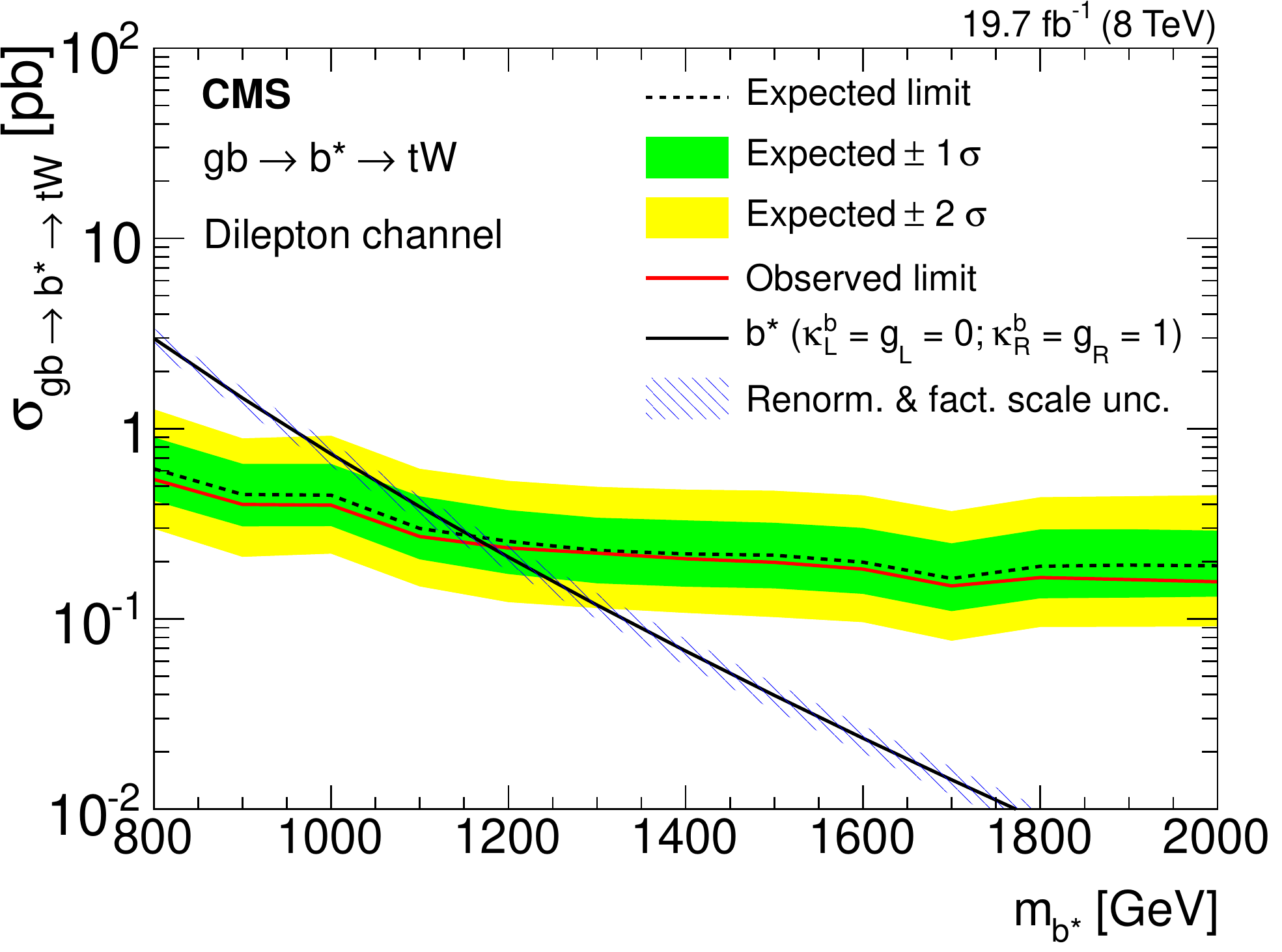}
\includegraphics[width=0.48\textwidth]{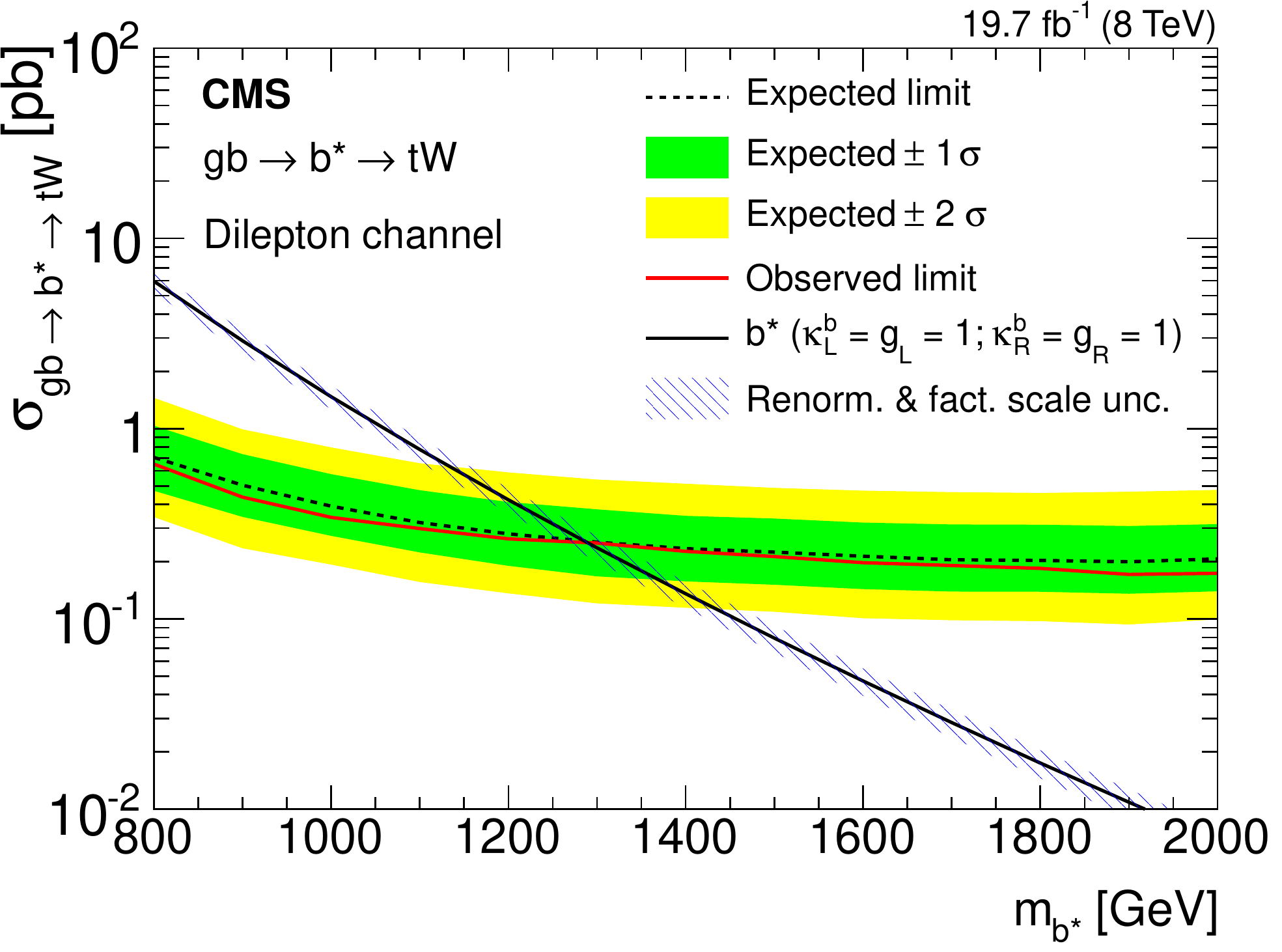}
\caption{The expected (dashed) and observed (solid) production cross section limits at 95\% CL for the dilepton channel as a function of $\bs$ quark mass for $ \Pg \PQb \to \bs \to \PQt \PW$. The theoretical cross section (solid line with hatched area) is also shown for comparison. The $1\sigma$ and $2\sigma$ uncertainties in the expected limit bands are shown. Limits for the left-handed, right-handed, and vector-like $\bs$ quark coupling hypotheses are shown in the top left, top right, and bottom plots, respectively.
           \label{dileplimitPlot}}
\end{figure}

A binned maximum likelihood fit to the $m_{\PQt \PW}$ distribution
is performed in both the all-hadronic and lepton+jets
channels, and to the $S_\mathrm{T}$ distribution in the dilepton channel
to extract the signal cross section. The observed distributions are consistent
with those from the background only prediction.
A Bayesian method~\cite[Ch. 38]{pdg} with a flat signal prior is
used within the \textsc{Theta} framework~\cite{theta} to set limits on
$\sigma_{\Pg \PQb \to \bs \to \PQt \PW}$.
The systematic uncertainties are accounted for as nuisance parameters,
and are integrated out using Bayesian marginalization.  Rate uncertainties
are modeled using log-normal priors. Uncertainties
varying as functions of the fitted variables are modeled using Gaussian priors,
and template morphing is employed to model the shape of these systematic uncertainties.
The limits on the cross section times branching fraction ($\sigma_{\Pg \PQb \to \bs \to \PQt \PW}$)
at 95\% confidence level (CL) are shown in
Figs.~\ref{fullHadlimitPlot},~\ref{lepjetlimitPlot},~and \ref{dileplimitPlot}
for the all-hadronic, lepton+jets, and dilepton channels, respectively.

\begin{figure}[phtb]
\centering
\includegraphics[width=0.48\textwidth]{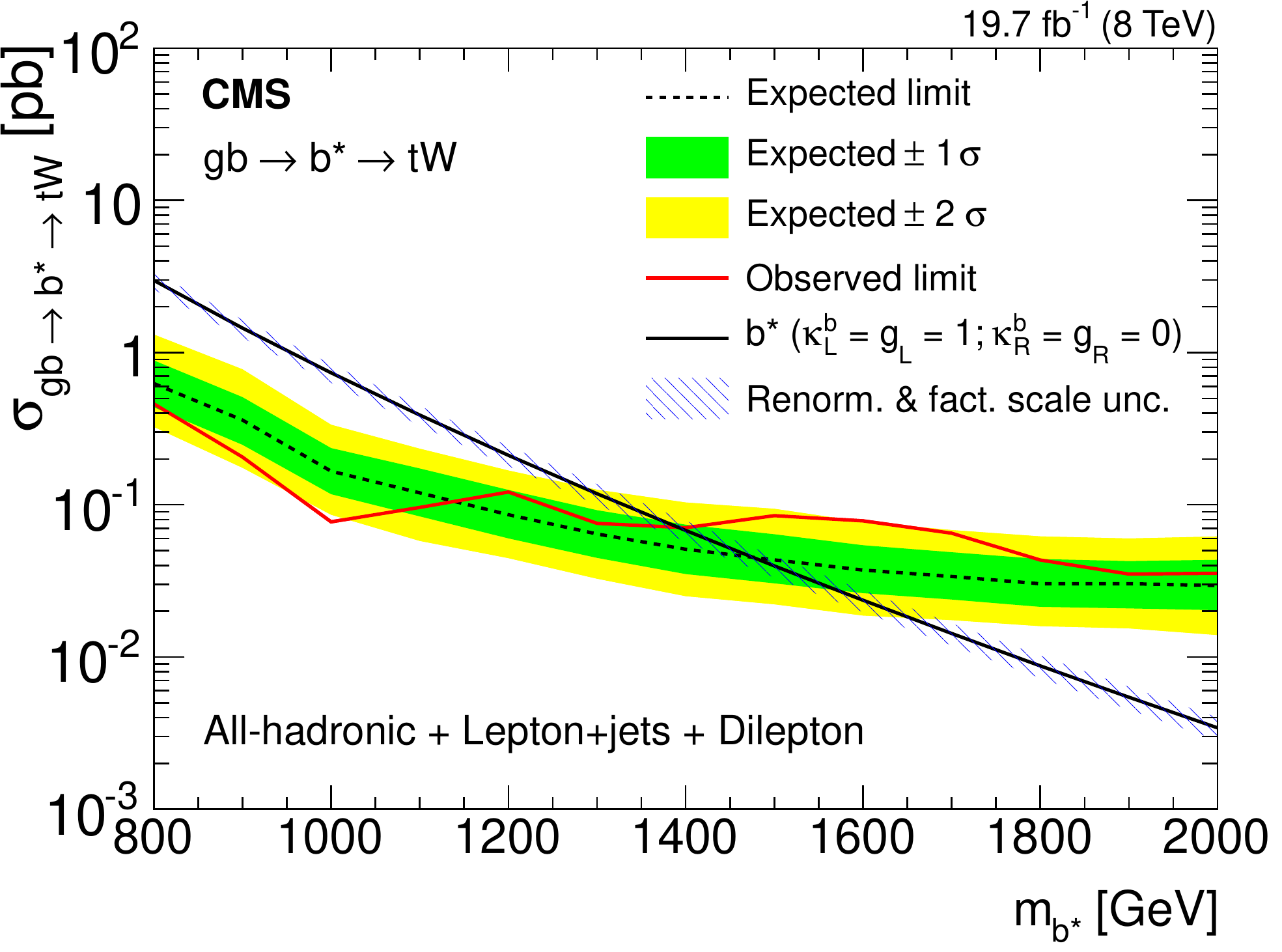}
\includegraphics[width=0.48\textwidth]{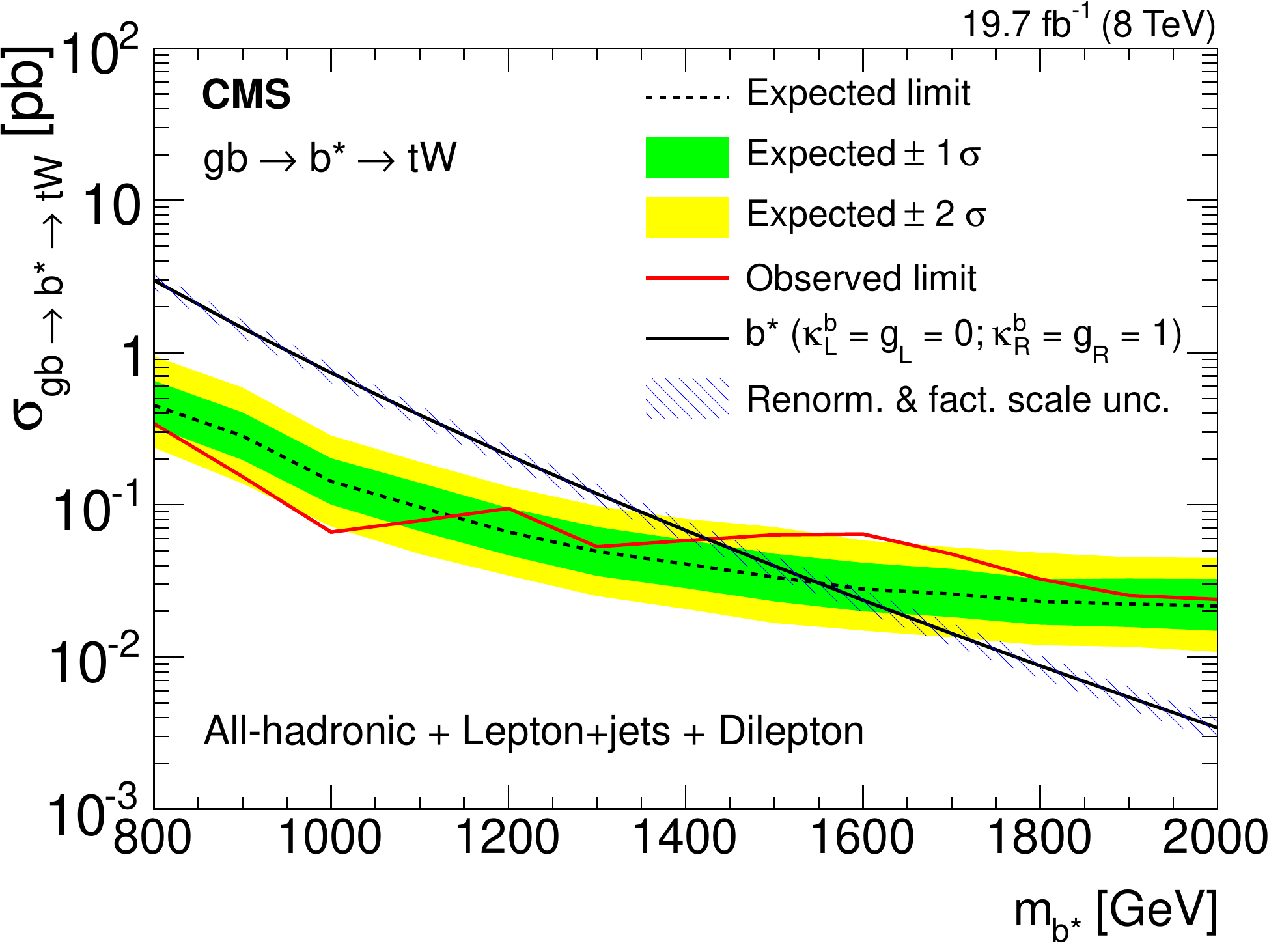}
\includegraphics[width=0.48\textwidth]{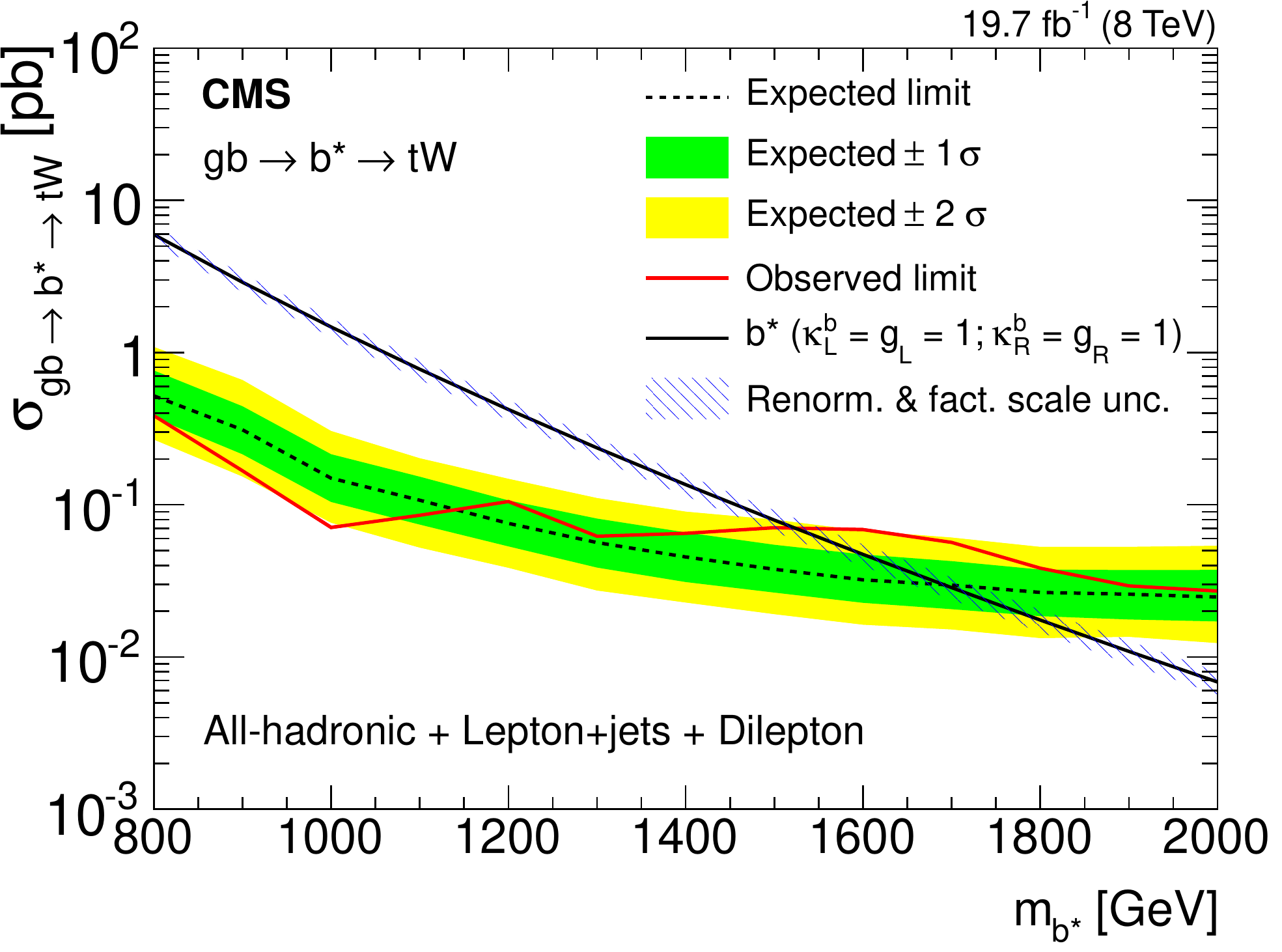}
\caption{ \normalsize The expected (dashed) and observed (solid) production cross section limits at 95\% CL for the combined all-hadronic, lepton+jets, and dilepton channels as a function of $\bs$ quark mass for $ \Pg \PQb \to \bs \to \PQt \PW$. The theoretical cross section (solid line with hatched area) is also shown for comparison. The $1\sigma$ and $2\sigma$ uncertainties in the expected limit bands are shown. Limits for the left-handed, right-handed, and vector-like $\bs$ quark coupling hypotheses are shown in the top left, top right, and bottom plots, respectively.
\label{figs:thetalimit}}
\end{figure}

To enhance the sensitivity of the measurement of the upper limit on the
$ \Pg \PQb \to \bs \to \PQt \PW$ production cross section,
the all-hadronic, lepton+jets, and dilepton channels are combined.
In forming the combination of separate channels, systematic uncertainties 
affecting both the shape and the event yield are taken into account.  
The procedure adopted is as follows: For each channel the shape of 
each distribution is determined and the normalization is set to 1. 
Then, for each bin ``i", an estimate is made of the systematic uncertainty 
$\sigma _{i}$ (not necessarily symmetric), which takes into account the 
contributions from all the sources affecting the shape. 
``Upper" and ``lower" distributions are then obtained, each normalized to 
unity, and used to estimate event yields in two limiting cases.
The systematic uncertainties are treated as being completely correlated
between bins of the distribution,
while the statistical uncertainties are treated as uncorrelated.
In the combination, the uncertainty sources due to jet energy scale,
jet energy resolution, b tagging scale factor, single top quark cross section, and integrated
luminosity are treated as correlated, and the remaining uncertainties are assumed to be
uncorrelated, as shown in Table~\ref{tab:corr_systematics}.
The limits are shown in Fig.~\ref{figs:thetalimit}.
The expected (observed) mass exclusion region at 95\% CL for the left-handed, right-handed,
and vector-like $\bs$ quark hypotheses is below 1480, 1560, and
1690\GeV (1390, 1430, and 1530\GeV), respectively as summarized in Table~\ref{limitTable}.

\begin{figure}[phtb]
\centering
\includegraphics[width=0.48\textwidth]{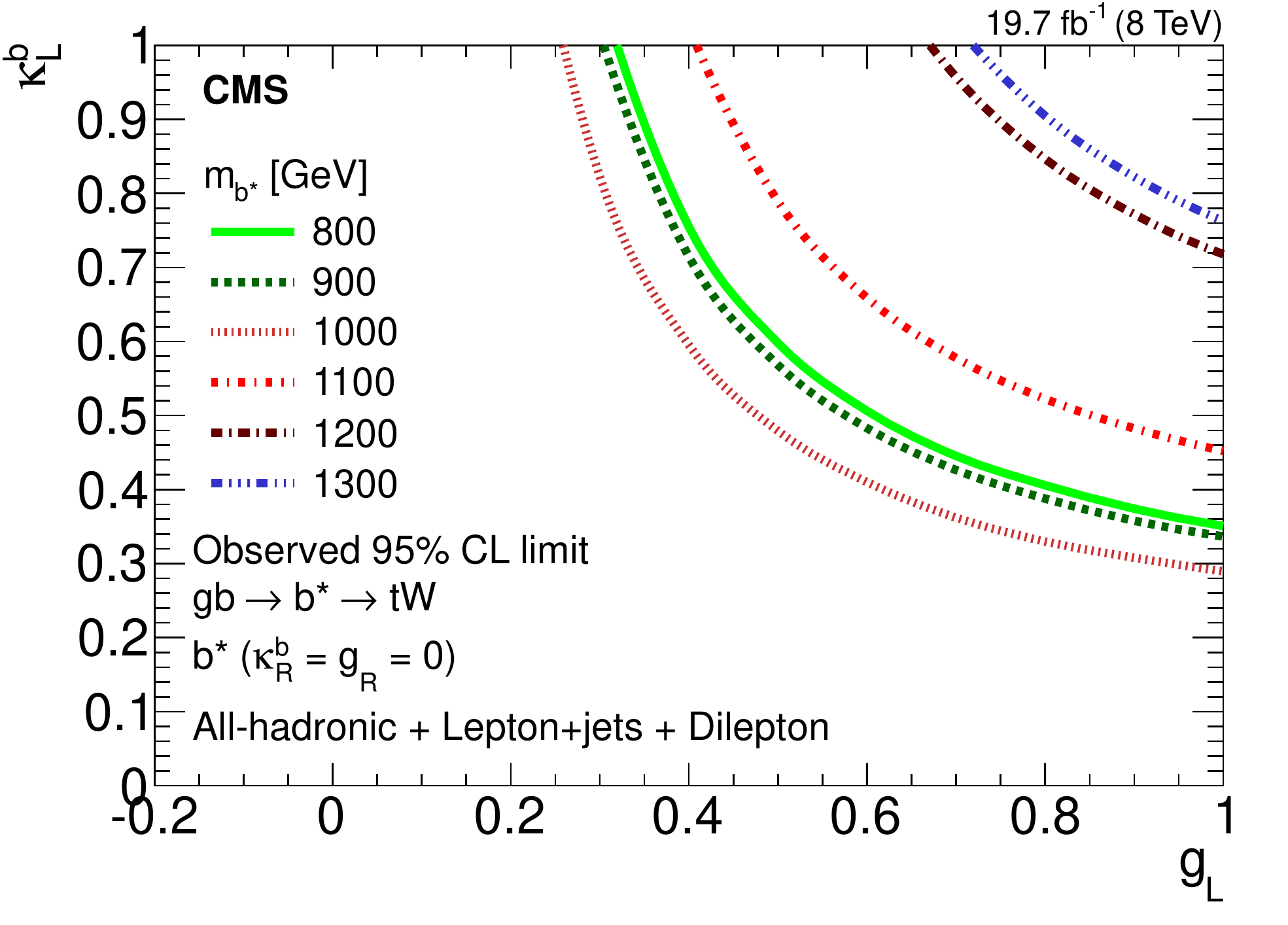}
\includegraphics[width=0.48\textwidth]{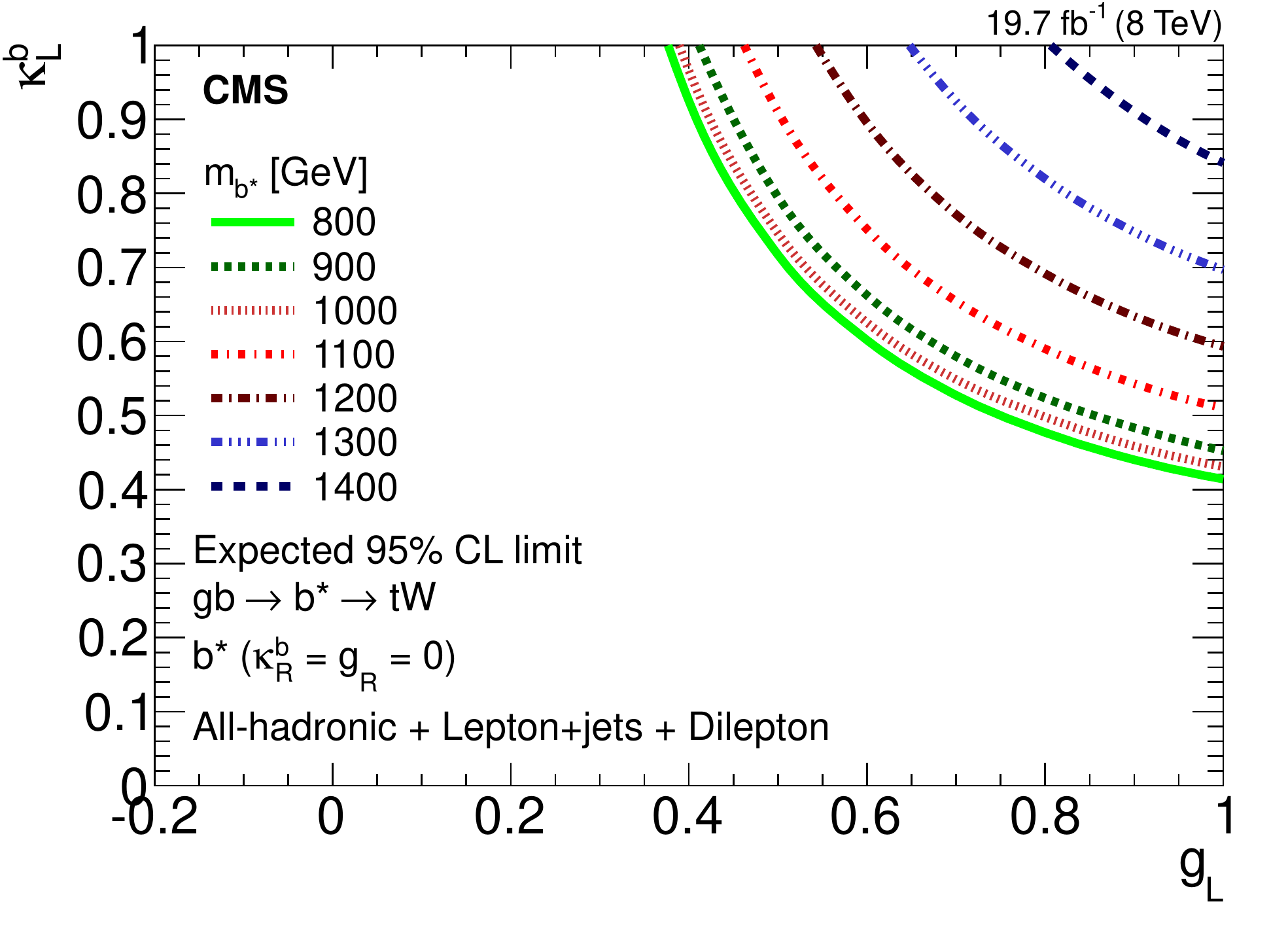}
\includegraphics[width=0.48\textwidth]{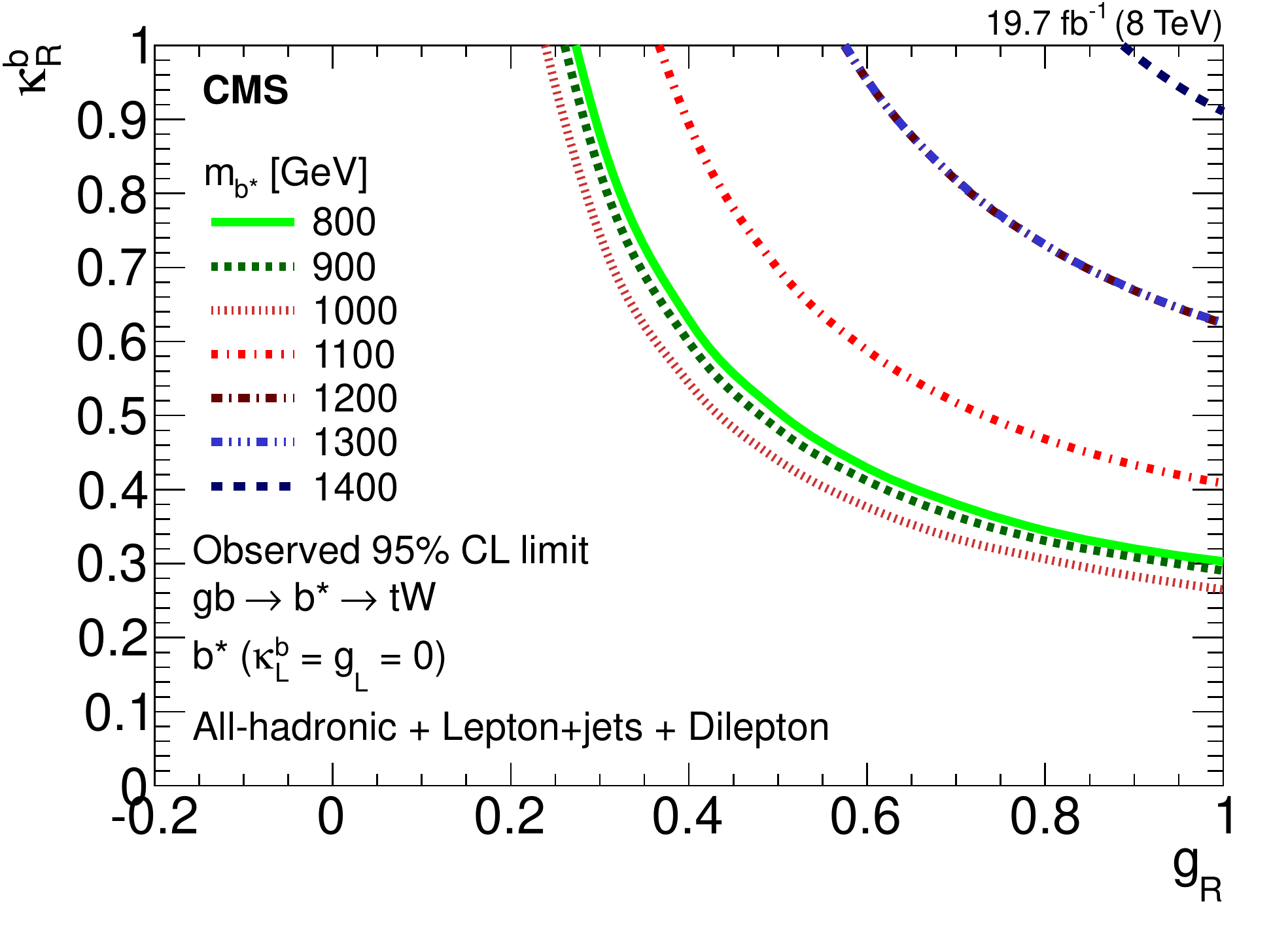}
\includegraphics[width=0.48\textwidth]{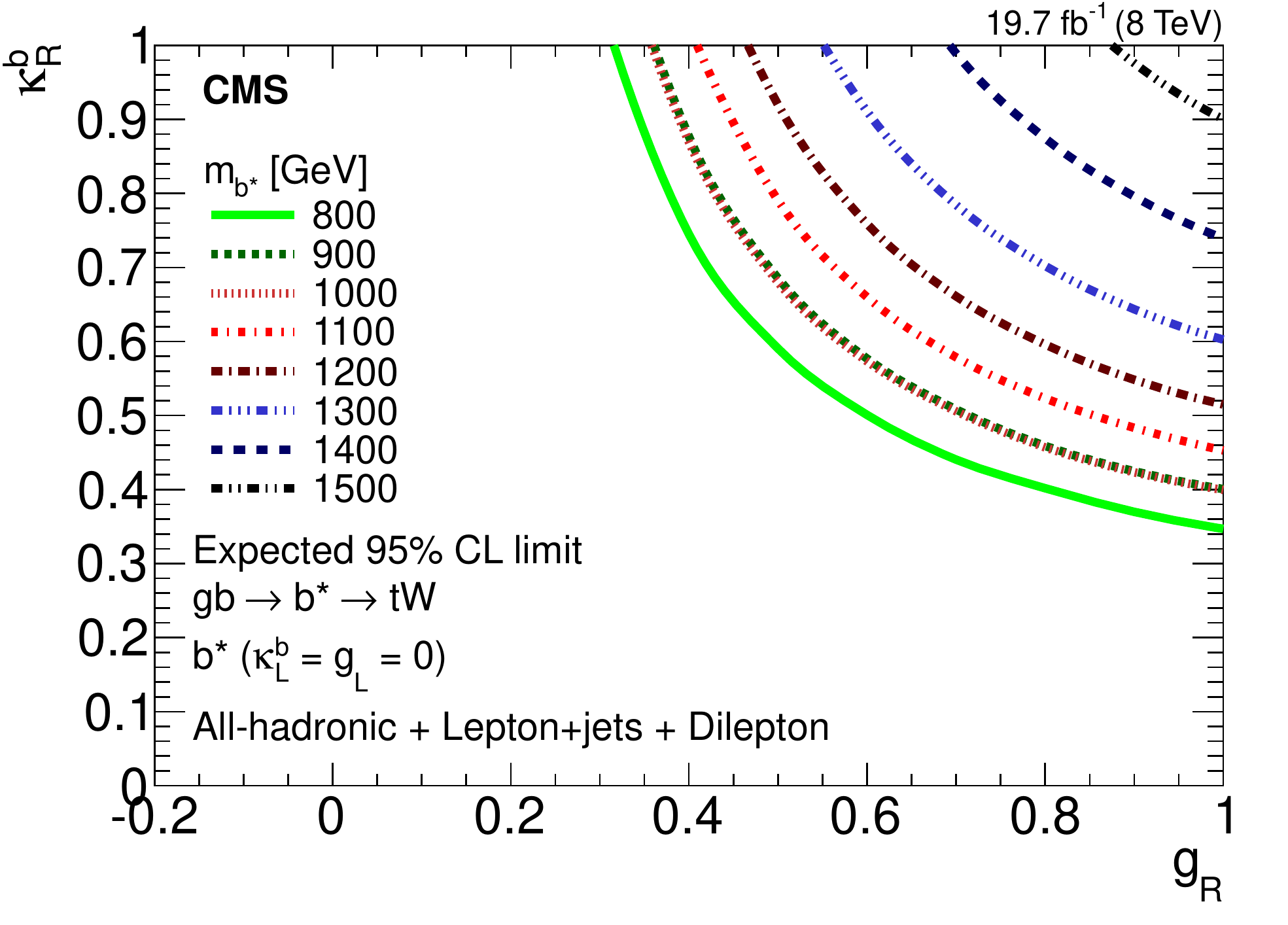}
\includegraphics[width=0.48\textwidth]{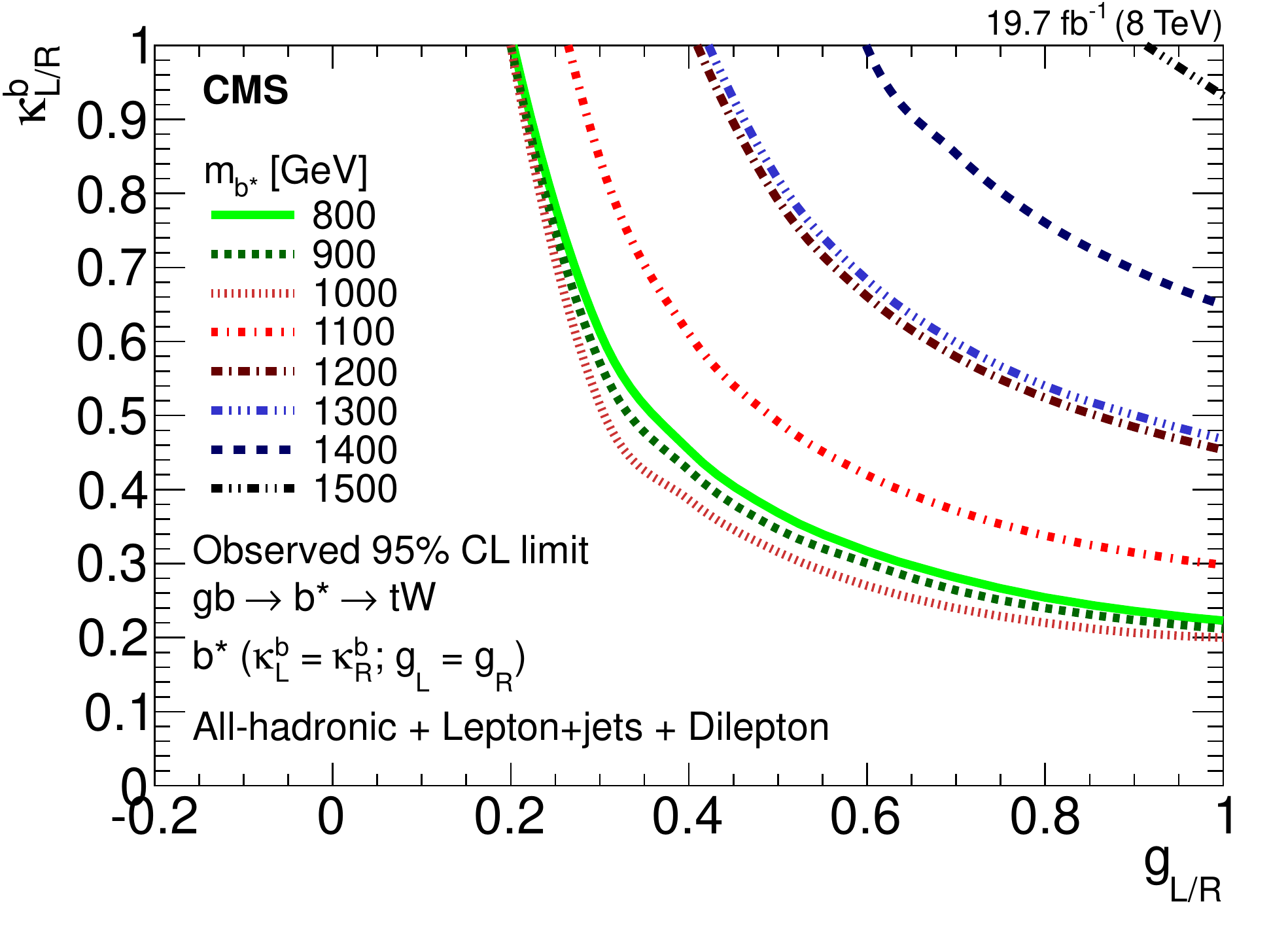}
\includegraphics[width=0.48\textwidth]{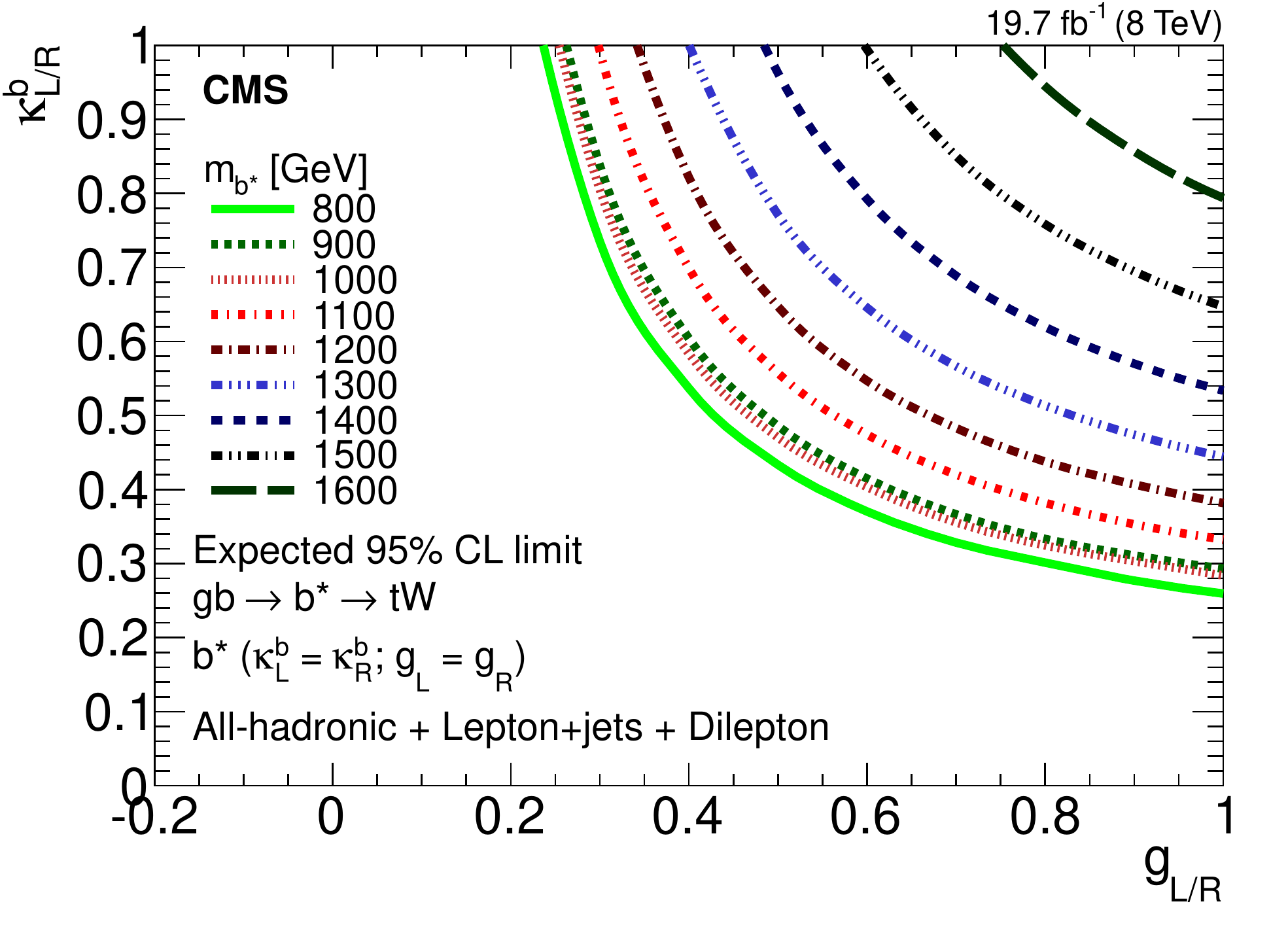}
\caption{ Contour plots showing the lower limits on various values of the $\bs$ quark mass, as a function of the couplings $\kappa$ and $g$. The left column shows the observed limits and the right column shows the expected limits. The limits for the left-handed, right-handed, and vector-like $\bs$ quark coupling hypotheses are shown in the top, middle, and bottom rows, respectively. The excluded regions are above and to the right of the curves.
\label{figs:thetalimit2}}
\end{figure}

The upper limits on the cross section times branching fraction
may be generalized as a function of the couplings $\kappa$ and $g$,
defined in equations~\ref{eq:productionL} and~\ref{eq:decayL}.
The results are shown in Fig.~\ref{figs:thetalimit2}.

\section{Summary}
\begin{table}[tb]
\centering
\caption{
The limit at 95\% CL, for the case of unit couplings, on $\bs$ quark
mass for the left-handed, right-handed, and vector-like coupling
hypotheses in the all-hadronic, lepton+jets dilepton, and
combined channels.
For each domain, two numbers linked with a dash indicate the excluded $\bs$ quark mass range, a single number
indicates the excluded lower $\bs$ quark mass limit.
}
\begin{tabular}{r|ccc}
\hline
&Left-handed&Right-handed&Vector-like\\
\hline
\multicolumn{4}{c}{All-hadronic channel} \\
\hline
Expected 95\% CL limit [\GeVns\!] & 890 - 1460 & 889 - 1520 & 842 - 1670 \\
Observed 95\% CL limit [\GeVns\!] & 858 - 1390 & 803 - 1430 & 1540 \\
\hline
\multicolumn{4}{c}{Lepton+jets channel} \\
\hline
Expected 95\% CL limit [\GeVns\!] & 935  & 985  & 1130 \\
Observed 95\% CL limit [\GeVns\!] & 1030 & 1070 & 1170 \\
\hline
\multicolumn{4}{c}{Dilepton channel} \\
\hline
Expected 95\% CL limit [\GeVns\!] & 1120 & 1170 & 1290 \\
Observed 95\% CL limit [\GeVns\!] & 1140 & 1180 & 1290 \\
\hline
\multicolumn{4}{c}{All-hadronic, lepton+jets, and dilepton channels combined} \\
\hline
Expected 95\% CL limit [\GeVns\!] &  1480 & 1560 & 1690 \\
Observed 95\% CL limit [\GeVns\!] &  1390 & 1430 & 1530 \\
\hline
\end{tabular}
\label{limitTable}
\end{table}

A search for a singly produced $\bs$ quark decaying
to $\PQt \PW$ in the all-hadronic, lepton+jets, and dilepton final states has been performed using proton-proton collisions recorded by the CMS
detector at $\sqrt{s}=8\TeV$, corresponding to an integrated luminosity of 19.7\fbinv.
No deviations that are inconsistent with standard model expectations are found
in the various spectra of variables used to search for the signal
in the three channels.
Upper limits are set at 95\% confidence level on the product of cross section
and branching fraction for the production of a $\bs$ quark that subsequently decays to $\PQt \PW$.
Excited bottom quarks are excluded with masses below 1390,
1430, and 1530\GeV for left-handed, right-handed, and vector-like $\bs$ quark couplings, respectively.
The mass limits are also extrapolated to the two dimensional $\kappa$-$g$ coupling plane.
These are the most stringent limits on the $\bs$ quark masses to date.

\section*{Acknowledgements}

\hyphenation{Bundes-ministerium Forschungs-gemeinschaft Forschungs-zentren} We congratulate our colleagues in the CERN accelerator departments for the excellent performance of the LHC and thank the technical and administrative staffs at CERN and at other CMS institutes for their contributions to the success of the CMS effort. In addition, we gratefully acknowledge the computing centres and personnel of the Worldwide LHC Computing Grid for delivering so effectively the computing infrastructure essential to our analyses. Finally, we acknowledge the enduring support for the construction and operation of the LHC and the CMS detector provided by the following funding agencies: the Austrian Federal Ministry of Science, Research and Economy and the Austrian Science Fund; the Belgian Fonds de la Recherche Scientifique, and Fonds voor Wetenschappelijk Onderzoek; the Brazilian Funding Agencies (CNPq, CAPES, FAPERJ, and FAPESP); the Bulgarian Ministry of Education and Science; CERN; the Chinese Academy of Sciences, Ministry of Science and Technology, and National Natural Science Foundation of China; the Colombian Funding Agency (COLCIENCIAS); the Croatian Ministry of Science, Education and Sport, and the Croatian Science Foundation; the Research Promotion Foundation, Cyprus; the Ministry of Education and Research, Estonian Research Council via IUT23-4 and IUT23-6 and European Regional Development Fund, Estonia; the Academy of Finland, Finnish Ministry of Education and Culture, and Helsinki Institute of Physics; the Institut National de Physique Nucl\'eaire et de Physique des Particules~/~CNRS, and Commissariat \`a l'\'Energie Atomique et aux \'Energies Alternatives~/~CEA, France; the Bundesministerium f\"ur Bildung und Forschung, Deutsche Forschungsgemeinschaft, and Helmholtz-Gemeinschaft Deutscher Forschungszentren, Germany; the General Secretariat for Research and Technology, Greece; the National Scientific Research Foundation, and National Innovation Office, Hungary; the Department of Atomic Energy and the Department of Science and Technology, India; the Institute for Studies in Theoretical Physics and Mathematics, Iran; the Science Foundation, Ireland; the Istituto Nazionale di Fisica Nucleare, Italy; the Ministry of Science, ICT and Future Planning, and National Research Foundation (NRF), Republic of Korea; the Lithuanian Academy of Sciences; the Ministry of Education, and University of Malaya (Malaysia); the Mexican Funding Agencies (CINVESTAV, CONACYT, SEP, and UASLP-FAI); the Ministry of Business, Innovation and Employment, New Zealand; the Pakistan Atomic Energy Commission; the Ministry of Science and Higher Education and the National Science Centre, Poland; the Funda\c{c}\~ao para a Ci\^encia e a Tecnologia, Portugal; JINR, Dubna; the Ministry of Education and Science of the Russian Federation, the Federal Agency of Atomic Energy of the Russian Federation, Russian Academy of Sciences, and the Russian Foundation for Basic Research; the Ministry of Education, Science and Technological Development of Serbia; the Secretar\'{\i}a de Estado de Investigaci\'on, Desarrollo e Innovaci\'on and Programa Consolider-Ingenio 2010, Spain; the Swiss Funding Agencies (ETH Board, ETH Zurich, PSI, SNF, UniZH, Canton Zurich, and SER); the Ministry of Science and Technology, Taipei; the Thailand Center of Excellence in Physics, the Institute for the Promotion of Teaching Science and Technology of Thailand, Special Task Force for Activating Research and the National Science and Technology Development Agency of Thailand; the Scientific and Technical Research Council of Turkey, and Turkish Atomic Energy Authority; the National Academy of Sciences of Ukraine, and State Fund for Fundamental Researches, Ukraine; the Science and Technology Facilities Council, UK; the US Department of Energy, and the US National Science Foundation.

Individuals have received support from the Marie-Curie programme and the European Research Council and EPLANET (European Union); the Leventis Foundation; the A. P. Sloan Foundation; the Alexander von Humboldt Foundation; the Belgian Federal Science Policy Office; the Fonds pour la Formation \`a la Recherche dans l'Industrie et dans l'Agriculture (FRIA-Belgium); the Agentschap voor Innovatie door Wetenschap en Technologie (IWT-Belgium); the Ministry of Education, Youth and Sports (MEYS) of the Czech Republic; the Council of Science and Industrial Research, India; the HOMING PLUS programme of the Foundation for Polish Science, cofinanced from European Union, Regional Development Fund; the OPUS programme of the National Science Center (Poland); the Compagnia di San Paolo (Torino); the Consorzio per la Fisica (Trieste); MIUR project 20108T4XTM (Italy); the Thalis and Aristeia programmes cofinanced by EU-ESF and the Greek NSRF; the National Priorities Research Program by Qatar National Research Fund; the Rachadapisek Sompot Fund for Postdoctoral Fellowship, Chulalongkorn University (Thailand); and the Welch Foundation, contract C-1845.

\bibliography{auto_generated}

\cleardoublepage \appendix\section{The CMS Collaboration \label{app:collab}}\begin{sloppypar}\hyphenpenalty=5000\widowpenalty=500\clubpenalty=5000\textbf{Yerevan Physics Institute,  Yerevan,  Armenia}\\*[0pt]
V.~Khachatryan, A.M.~Sirunyan, A.~Tumasyan
\vskip\cmsinstskip
\textbf{Institut f\"{u}r Hochenergiephysik der OeAW,  Wien,  Austria}\\*[0pt]
W.~Adam, E.~Asilar, T.~Bergauer, J.~Brandstetter, E.~Brondolin, M.~Dragicevic, J.~Er\"{o}, M.~Flechl, M.~Friedl, R.~Fr\"{u}hwirth\cmsAuthorMark{1}, V.M.~Ghete, C.~Hartl, N.~H\"{o}rmann, J.~Hrubec, M.~Jeitler\cmsAuthorMark{1}, V.~Kn\"{u}nz, A.~K\"{o}nig, M.~Krammer\cmsAuthorMark{1}, I.~Kr\"{a}tschmer, D.~Liko, T.~Matsushita, I.~Mikulec, D.~Rabady\cmsAuthorMark{2}, B.~Rahbaran, H.~Rohringer, J.~Schieck\cmsAuthorMark{1}, R.~Sch\"{o}fbeck, J.~Strauss, W.~Treberer-Treberspurg, W.~Waltenberger, C.-E.~Wulz\cmsAuthorMark{1}
% ...
\vskip\cmsinstskip
% ...
1:~~Also at Vienna University of Technology, Vienna, Austria\\
2:~~Also at CERN, European Organization for Nuclear Research, Geneva, Switzerland\\
% ...

\end{sloppypar}
\end{document}